\documentclass[twocolumn, times]{aastex631}
\usepackage{gensymb}
\usepackage{amssymb}
\usepackage{amsmath}
\usepackage{hepnames}

\usepackage{wasysym}
\usepackage[utf8]{inputenc}
\usepackage[T1]{fontenc}
\usepackage{fancyvrb}

\newcommand{\besselj}[1]{J_{{#1}}}
\newcommand{\cond}{\mathrm{S}~\mathrm{m}^{-1}}
\newcommand{\rocean}{R_\mathrm{ocean}}
\newcommand{\rmantle}{R_\mathrm{mantle}}
\newcommand{\rmoon}{R_\mathrm{moon}}
\newcommand{\reuropa}{R_\mathrm{Eur}}
\newcommand{\unitvec}[1]{\boldsymbol{\hat{{#1}}}}
\newcommand{\oschat}{\unitvec{e}}

\shorttitle{Interior structure of icy moons from Bayesian inversion of magnetometry}
\shortauthors{Biersteker et al.}

\begin{document}

\title{Revealing the interior structure of icy moons with a Bayesian approach to magnetic induction measurements}

\correspondingauthor{John B. Biersteker}
\email{jo22395@mit.edu}

\author[0000-0001-5243-241X]{John B. Biersteker}
\affiliation{Department of Earth, Atmospheric, and Planetary Sciences, Massachusetts Institute of Technology, 77 Massachusetts Avenue, Cambridge, MA 02139-4307, USA}

\author{Benjamin P. Weiss}
\affiliation{Department of Earth, Atmospheric, and Planetary Sciences, Massachusetts Institute of Technology, 77 Massachusetts Avenue, Cambridge, MA 02139-4307, USA}

\author{Corey J. Cochrane}
\affiliation{Jet Propulsion Laboratory, California Institute of Technology, Pasadena, CA, USA}

\author{Camilla D. K. Harris}
\affiliation{Jet Propulsion Laboratory, California Institute of Technology, Pasadena, CA, USA}
\affiliation{Department of Climate and Space Sciences and Engineering, University of Michigan, Ann Arbor, MI, USA}

\author{Xianzhe Jia}
\affiliation{Department of Climate and Space Sciences and Engineering, University of Michigan, Ann Arbor, MI, USA}

\author{Krishan K. Khurana}
\affiliation{Department of Earth, Planetary and Space Sciences and Institute of Geophysics and Planetary Physics, University of California, Los Angeles, CA, USA}

\author{Jiang Liu}
\affiliation{Department of Earth, Planetary and Space Sciences and Institute of Geophysics and Planetary Physics, University of California, Los Angeles, CA, USA}

\author{Neil Murphy}
\affiliation{Jet Propulsion Laboratory, California Institute of Technology, Pasadena, CA, USA}

\author{Carol A. Raymond}
\affiliation{Jet Propulsion Laboratory, California Institute of Technology, Pasadena, CA, USA}

% Abstract
\begin{abstract}
Some icy moons and small bodies in the solar system are believed to host subsurface liquid water oceans. 
The interaction of these saline, electrically conductive oceans with time-varying external magnetic fields generates induced magnetic fields. Magnetometry observations of these induced fields in turn enable the detection and characterization of these oceans.
We present a framework for characterizing the interiors of icy moons using multi-frequency induction and Bayesian inference applied to magnetometry measurements anticipated from the upcoming Europa Clipper mission. Using simulated data from the Europa Clipper Magnetometer (ECM), our approach can accurately retrieve  a wide range of plausible internal structures for Europa. In particular, the ocean conductivity is recovered to within ${\pm}50\%$ for all internal structure scenarios considered and the ocean thickness can be retrieved to within ${\pm}25~\mathrm{km}$ for five out of seven scenarios. Characterization of the ice shell thickness to ${\pm}50\%$ is possible for six of seven scenarios. Our recovery of the ice shell thickness is highly contingent on accurate modeling of magnetic fields arising from the interaction of Europa with the ambient magnetospheric plasma, while the ocean thickness is more modestly affected and the ocean conductivity retrieval is largely unchanged. Furthermore, we find that the addition of a priori constraints (e.g., static gravity measurements) can yield improved ocean characterization compared to magnetometry alone, suggesting that  multi-instrument techniques can play a key role in revealing the interiors of Europa and other ocean worlds.
\end{abstract}

\keywords{Europa (2189), Galilean satellites (627), Magnetic fields (994), Markov chain Monte Carlo (1889), Planetary interior (1248)}

%%%%% Section: Introduction
\section{Introduction}
\label{sec: introduction}
Spacecraft exploration of the solar system has revealed a diverse collection of icy satellites and dwarf planets, many of which may be ocean worlds harboring large bodies of liquid water beneath their frozen surfaces \citep{2016JGRE..121.1378N}. Detecting and characterizing these oceans are important objectives of future spacecraft missions. Of particular interest is establishing whether, in addition to liquid water, these environments possess the chemical building blocks and energy sources needed to sustain life \citep{2020SSRv..216...95H}. The upcoming Europa Clipper mission to the Jupiter system aims to answer these questions to assess the habitability of Europa's global subsurface ocean \citep{2020NatCo..11.1311H}.

The presence of an ocean on Europa was established by the Galileo spacecraft \citep{1998Natur.395..777K, 1998JGR...10319843N,2000Sci...289.1340K}, but the structure of the ice and liquid water layers and the ocean's salinity remain poorly constrained \citep{2005Icar..177..397B, 2007Icar..189..424H, 2009euro.book..571K}. The thickness of the overlying ice shell controls the mechanism and rate of delivery of radiolytically produced oxidants from Europa's surface to the ocean, and the depth of the seafloor affects the degree of water-rock reactions and production rate of hydrogen \citep{2007AsBio...7.1006H, 2016GeoRL..43.4871V, 2020SSRv..216...80S}. The global-scale thicknesses of the ice and ocean are therefore key astrobiological parameters and observational targets for Europa Clipper.

Spacecraft magnetometry offers a powerful probe of the interiors of icy satellites through the detection of induced magnetic fields. These fields are generated by the interaction of electrically conductive layers in satellite interiors with the time-varying magnetospheric field of their host planet. The presence of a subsurface ocean on Europa was first confirmed by the detection of a time-varying dipolar field originating from Europa that co-varied with the changing Jovian background field. This observation is best explained by a global, near-surface conducting layer, most probably a salty liquid water ocean \citep{2000Sci...289.1340K}. 
Because the salinity and thickness of the ocean and the thickness of the overlying ice shell all affect the induced magnetic response, spacecraft magnetometry has the potential to recover critical information about Europa's internal structure and habitability \citep[e.g.,][]{2000Icar..147..329Z, 2011Icar..214..477S}.

However, because of the dependency of the induction response on these three parameters, analysis of the Galileo magnetometry data has not yielded a unique internal structure for Europa. Specifically, the number and timing of Galileo flybys of Europa limit previous studies to using only the induction response to magnetic variation at the synodic period (i.e., the time required for Europa to return to the same Jovian longitude, $11.2 ~\mathrm{hr}$) \citep{2000Icar..147..329Z, 2004JGRE..109.5006S, 2007Icar..192...41S, 2007Icar..189..424H}. This results in a degeneracy between the internal structure parameters (see Section \ref{sec: induction}) that hampers direct recovery of the ocean parameters.
Instead, prior studies identified bounds on the range of plausible internal structures. \citet{2000Icar..147..329Z} found that an ocean with a conductivity ${>}0.06~ \cond$ located ${\lesssim}200~\mathrm{km}$ from the surface matches the data given the uncertainties associated with unmodeled magnetospheric plasma processes. Subsequent analysis \citep{2004JGRE..109.5006S, 2007Icar..192...41S} favors an ocean nearer the surface (${\lesssim}20~\mathrm{km}$) with a conductivity ${>}0.5~\cond$ and thickness ${\lesssim}100~\mathrm{km}$. Incorporating physically realistic limits on the conductivity for $\mathrm{MgSO_4}$ and $\mathrm{NaCl}$, \citet{2007Icar..189..424H} also favor a thin ice shell of ${<}15~\mathrm{km}$. A key goal of the Europa Clipper Magnetometer (ECM) investigation is to break the degeneracies between ocean conductivity, ocean thickness, and ice thickness by measuring the magnetic response of Europa at multiple frequencies \citep{2015AGUFM.P13E..08R}. In conjunction with measurements from other instruments, these data will allow Europa Clipper to achieve its science objectives to constrain the globally-averaged ice shell and ocean thickness, as well as the ocean's salinity, to within $\pm50\%$ \citep{9172447}.

We present a new Bayesian inference technique for recovering the internal structures of icy satellites from multi-frequency magnetic induction measurements. The particular advantages of this technique are (i) that it provides direct estimates of the ocean parameters with (ii) robust uncertainties, including quantification of parameter degeneracies, and (iii) naturally incorporates prior constraints. We apply this technique to simulated ECM data from the more than 40 Europa flybys similar to those Europa Clipper will perform to demonstrate the feasibility of recovering unique ocean structures. In Section \ref{sec: induction}, we review the use of magnetic induction for studying planetary interiors with a focus on the Jupiter-Europa system. In Section \ref{sec: clipper magnet}, we describe the simulated ECM data. We introduce our Bayesian inversion technique in Section \ref{sec: bayesian retrieval} and present retrievals of simulated interiors with our technique in Section \ref{sec: results}. We close in Section \ref{sec: discussion} with a discussion of the implications for ECM and other future applications.

%%%%% Section: Induction physics
\section{Magnetic induction}
\label{sec: induction}
\subsection{Conducting Shell Model}
\begin{figure}
	\centering
	\includegraphics[width=0.5\textwidth]{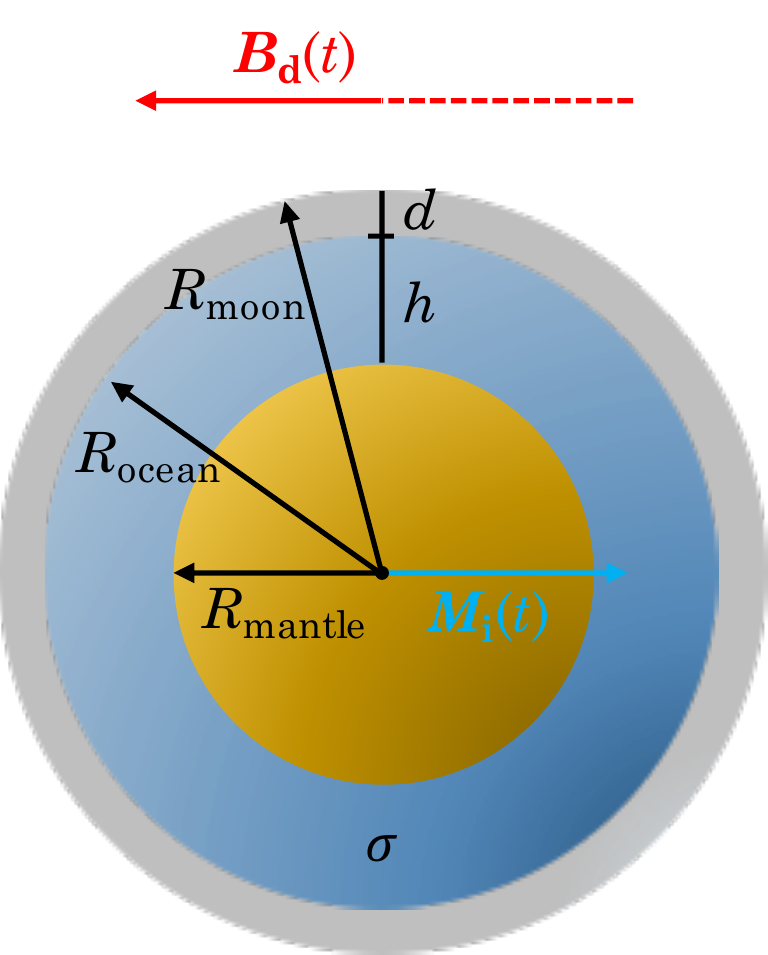}
	\caption{\small Three-layer internal structure model of Europa. The induction response is determined by the ice shell thickness ($d$), ocean thickness ($h$), and ocean conductivity ($\sigma$). Also shown is a schematic of an oscillating driving field ($\boldsymbol{B_\mathrm{d}}{(t)}$, red, Equation [\ref{eq: driving}]) and the induced magnetic moment ($\boldsymbol{M_\mathrm{i}}{(t)}$, blue, Equation [\ref{eq: induced moment}]) generated in response, which points opposite the time-variable field. After \citet{2000Icar..147..329Z}.}
	\label{fig: interior structure}
\end{figure}

Time-varying magnetic fields inside a conductor generate electric currents, which in turn give rise to an induced magnetic field. Measuring these induced fields provides a probe of the electrical conductivity structure of planetary bodies \citep[for a review, see][]{2010SSRv..152..391S}. We assume a three-layer spherically symmetric internal structure for Europa after \citet{2000Icar..147..329Z}, consisting of a single conductive layer between a negligibly conductive ice shell and rocky mantle (Figure \ref{fig: interior structure}). The ocean layer is assumed to have a uniform conductivity, $\sigma$, with inner and outer radii $\rmantle$ and $\rocean$, respectively. The radius of the moon is $\rmoon = R_\mathrm{Europa} = 1560.8~\mathrm{km}$. For convenience, we also define $d = \rmoon - \rocean$ as the ice shell thickness and $h = \rocean - \rmantle$ as the ocean thickness. The magnetic response predicted by this idealized model differs somewhat from that predicted by more sophisticated internal structure models \citep[e.g.,][]{2021Icar..35414020S,2021JGRE..12606418V}, but the three-layer model remains suitable for our goal of assessing the ability of ECM to recover the global properties of the ocean. We address these differences in more detail in Section \ref{sec: discussion}.

Under the assumption of zero displacement current, the magnetic field, $\boldsymbol{B}$, obeys the diffusion equation:
\begin{align}
\nabla^2{\boldsymbol{B}} = \mathrm{\mu}_0 \sigma \frac{\partial{\boldsymbol{B}}}{\partial{t}} \text{,}
\end{align}
where we take the magnetic permeability to be equal to the vacuum permeability, $\mathrm{\mu}_0$, everywhere. A sinusoidally-varying driving field ($\boldsymbol{B_\mathrm{d}}$) oscillating in a direction given by the unit vector $\oschat$, can be represented as the real component of
\begin{align}
\label{eq: driving}
\boldsymbol{B_\mathrm{d}} = B_\mathrm{d} e^{- i \omega t} \oschat \text{,}
\end{align}
where $\omega$ is the angular frequency of the oscillation. Finally, following \citet{2000Icar..147..329Z}, we assume that the Jovian magnetic field is spatially uniform on the scale of $\rmoon$.

The solution for the induced field generated by a conducting spherical shell in an oscillating, spatially uniform driving field is derived in \citet{osti_6997191}. The resulting field outside the conductor is dipolar with a time-variable moment aligned with the oscillation axis:
\begin{align}
\label{eq: induced moment}
\boldsymbol{M_\mathrm{i}} = - \frac{2 \mathrm{\pi}}{\mathrm{\mu}_0} A e^{i \phi} \boldsymbol{B_\mathrm{d}} \rmoon^3 \text{,}
\end{align}
where $A \in [0, 1]$ is the amplitude of the induction response and $\phi \in [0, \mathrm{\pi} / 2]$ represents a phase lag between the driving field and the induced field. The complex amplitude is
\begin{align}
\label{eq: complex amplitude}
A e^{i \phi}  & = \left( \frac{\rocean}{\rmoon} \right)^3 \frac{\xi \besselj{5/2}(\rocean k) - \besselj{-5/2}(\rocean k)}{\xi \besselj{1/2}(\rocean k) - \besselj{-1/2}(\rocean k)} \text{,}
\end{align}
where
\begin{align}
\xi &= \frac{\rmantle k \besselj{-5/2}(\rmantle k)}{3 \besselj{3/2}(\rmantle k ) - \rmantle k \besselj{1/2}(\rmantle k)} \text{,}
\end{align}
$\besselj{\nu}(z)$ is the Bessel function of the first kind with order $\nu$ and argument $z$, and $k = (1 + i) \sqrt{\mathrm{\mu}_0 \sigma \omega /2}$ is the complex wave vector \citep{osti_6997191}. The induced magnetic field at a position relative to the center of the moon $\boldsymbol{r} = r \unitvec{r}$ can then be written
\begin{align}
\label{eq: induced field}
\boldsymbol{B_\mathrm{i}} = - A e^{-i (\omega t - \phi)} B_\mathrm{d} \frac{\rmoon^3}{ r^3 } \frac{ 3 \left( \unitvec{r} \boldsymbol{\cdot} \oschat \right) \unitvec{r} - \oschat}{2} \text{.}
\end{align}

The details of this model applied to the case of icy satellites are explored in \citet{2000Icar..147..329Z}. Here we briefly recapitulate some salient features. In the limit of a perfectly conducting layer ($\sigma \rightarrow \infty$), the complex amplitude $A e^{i \phi} \rightarrow (\rocean/\rmoon)^3$ and there is no phase lag between the induced and driving fields ($\phi \rightarrow 0$). This case provides the upper bound for the amplitude of the induced field, $B_\mathrm{i, perf} = B_\mathrm{d}$ at the pole of the induced dipole. At the surface of the perfect conductor ($r = \rocean$), the combined induced and driving field is
\begin{align}
\boldsymbol{B_\mathrm{d}} + \boldsymbol{B_{\mathrm{i}, \mathrm{perf}}} = B_\mathrm{d} e^{-i \omega t} \frac{3}{2} \left[ \oschat -  \left( \unitvec{r} \boldsymbol{\cdot} \oschat \right) \unitvec{r} \right] \text{,}
\end{align}
which, as can be seen by the vector expression, is zero at the induction poles ($\unitvec{r} = \pm \oschat$) and everywhere tangent to the conductor surface. 

Finally, this idealized case provides an example of the degeneracy in recovering ocean parameters from a single observation of the induction response. Measuring the induction amplitude and phase in this scenario constrains $\rocean$, and therefore $d$, the ice shell thickness, but cannot probe $\rmantle$ or the ocean thickness, $h$.  For physical conductors, the amplitude of the induction response decreases and the response begins to lag the driving field, providing a probe into the moon's internal structure, but degeneracies remain unless the induction response can be measured at multiple frequencies.

\subsection{Jovian Field at Europa}
\label{sec: jovian field}
The orbital motion of Europa, rotation of Jupiter, and the dynamic Jovian magnetosphere combine to create a complex time-variable field at Europa that can drive induction \citep{2009euro.book..571K, 2011Icar..214..477S}. 
This driving field and the induced response can be decomposed into a summation of sinusoidal oscillations at discrete frequencies along the axes of the coordinate system, each represented by Equation \eqref{eq: driving}. 

\begin{figure}
	\centering
	\includegraphics[width=0.5\textwidth]{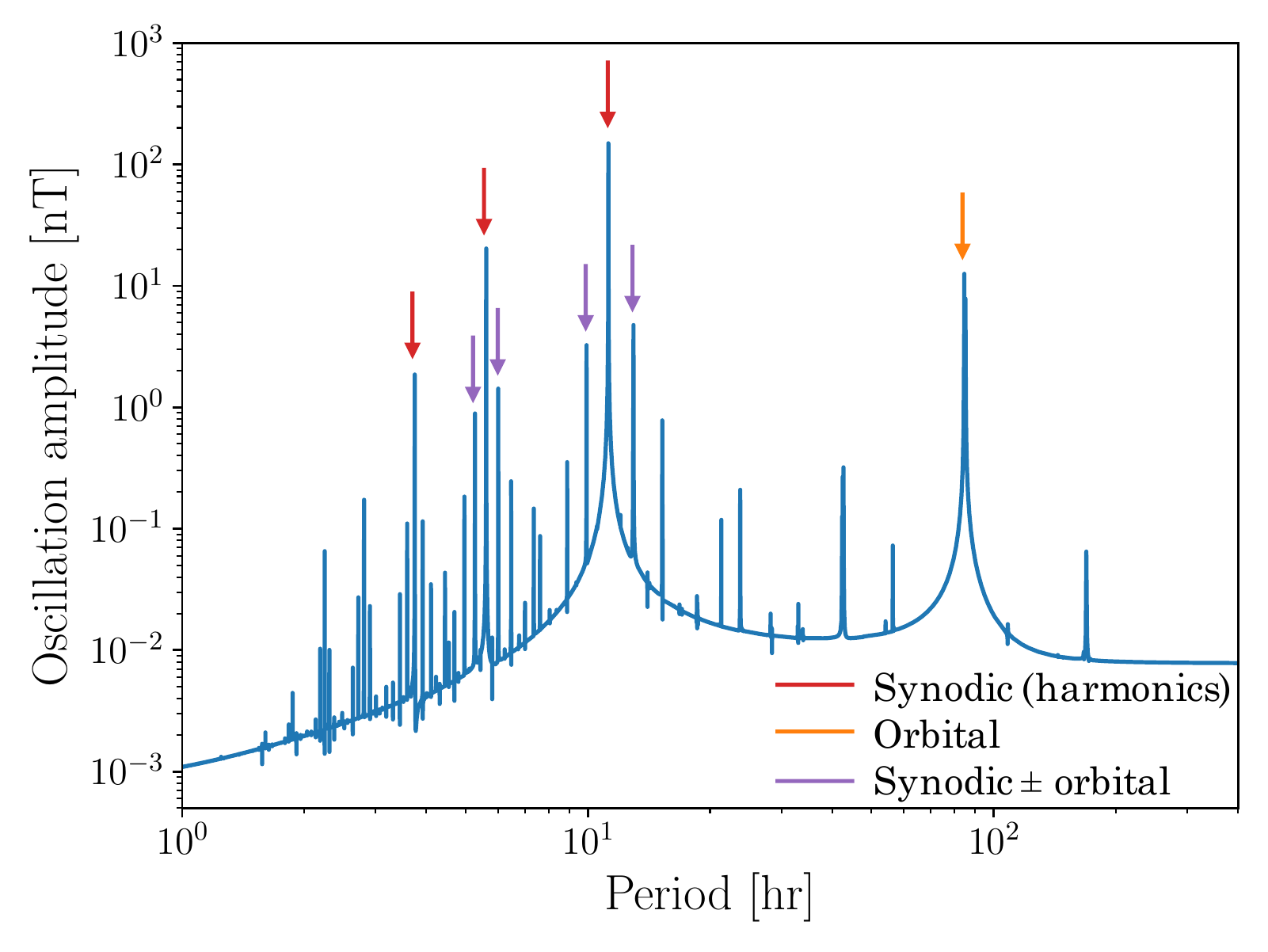}
	\caption{\small Simulated spectrum of the ambient Jovian magnetic field variation at Europa. Peaks with a magnitude ${>}1~\mathrm{nT}$ are indicated with arrows showing their origin, with red for the synodic period and its harmonics, orange for the orbital motion, and purple for beats between synodic and orbital terms. The peak near $85~\mathrm{hr}$ is bimodal (see Figure \ref{fig: zoomed Jovian field FFT at Europa}).}
	\label{fig: Jovian field FFT at Europa}
\end{figure}

\begin{figure}
	\centering
	\includegraphics[width=0.5\textwidth]{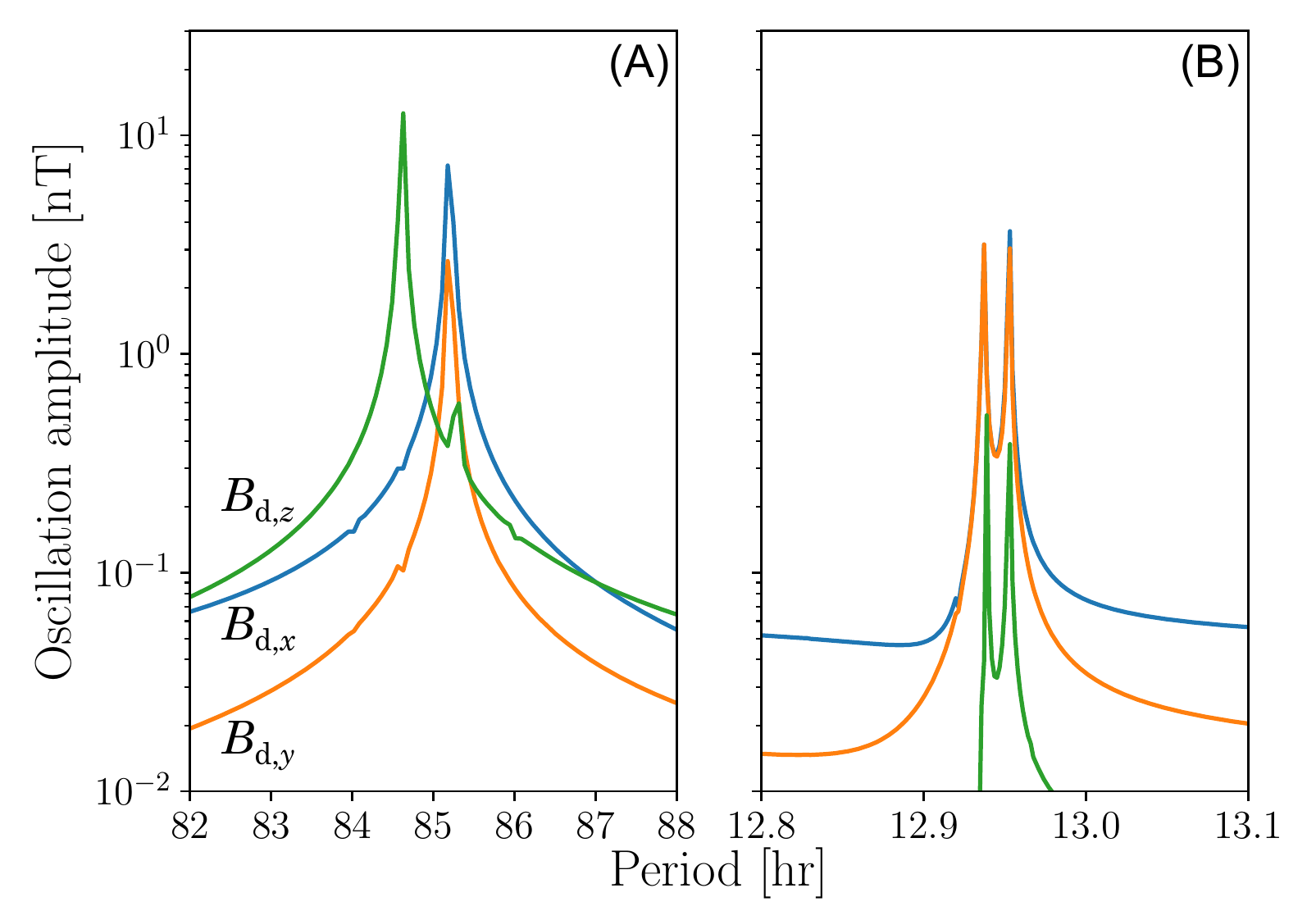}
	\caption{\small Simulated spectrum of the ambient magnetic field variation at Europa along each axis of the IAU Europa body-fixed coordinate system, zoomed to highlight bimodal features. $B_{\mathrm{d},x}$ is shown in blue, $B_{\mathrm{d},y}$ in orange, and $B_{\mathrm{d},z}$, in green. \textbf{(A)} Bimodal feature near the orbital period caused by apsidal and nodal precession. \textbf{(B)} Bimodal feature at the lower frequency sideband of the synodic frequency caused by different modulations of the synodic oscillation.
	}
	\label{fig: zoomed Jovian field FFT at Europa}
\end{figure}

To explore the driving field at Europa, we simulated a time series of the ambient magnetic field at Europa over a period of $12 ~\mathrm{yr}$, roughly one Jupiter orbit around the Sun, with a cadence of $20~\mathrm{min}$ using a version of the global Jovian magnetosphere model of \citep[][]{2004cosp...35.2073K}, which provides a good match to the Galileo data.
This model includes the VIP4 Jovian internal field model \citep{1998JGR...10311929C}, the warped and delayed current sheet model provided by \citet{2005JGRA..110.7227K}, the shielding field of the magnetopause, and incorporates penetration of the magnetosphere by the interplanetary magnetic field. The internal field and current sheet dominate the magnetic field variation at Europa, while the inclusion of the magnetospheric structure adds additional variability associated with Europa's local time.

To describe the motions of Europa and Jupiter we used SPICE kernels \citep{2018P&SS..150....9A} provided by NASA's Navigation and Ancillary Information Facility (NAIF).
We calculated the oscillation for each component of the International Astronomical Union (IAU) body-fixed coordinate system for Europa \citep{2018CeMDA.130...22A}, where the $Z$-axis points along Europa's rotation axis, which is assumed to be synchronous and normal to the mean orbital plane, the $X$-axis points towards Jupiter, and the $Y$-axis completes the right-handed triad.
After taking a fast Fourier transform of the simulated time series, the resulting spectrum shows eleven peaks each with magnitude $(B_{\mathrm{d},x}^2 + B_{\mathrm{d},y}^2 + B_{\mathrm{d},z}^2)^{1/2} > 1 ~\mathrm{nT}$: three from the synodic period and harmonics, two from Europa's orbital motion, and six beat frequencies of these two signals (Figure \ref{fig: Jovian field FFT at Europa}).

A comprehensive breakdown of the origin of the different primary frequencies is provided in \citet{2011Icar..214..477S}. Variation at the synodic period ($11.23~\mathrm{hr}$) is primarily caused by the wobbling of Jupiter's dipole axis due to its $9.6 \degree$ tilt relative to Jupiter's spin axis \citep[e.g.,][]{2009euro.book..571K}, with a contribution from the current sheet \citep{2011Icar..214..477S}. The non-dipolar part of the internal Jovian field and current sheet are responsible for the harmonics of the synodic period at $5.62$ and $3.74~\mathrm{hr}$ \citep{2011Icar..214..477S}. Europa's orbital motion ($P_\mathrm{orb} = 85.23~\mathrm{hr}$) introduces variation due to its orbital inclination, which modifies the tilted dipole geometry, and eccentricity, which modulates the field strength as experienced at Europa. Additional variation at the orbital period comes from the day-night asymmetry in the Jovian magnetosphere, which is compressed by the solar wind on the dayside \citep{2001JGR...10625999K, 2009euro.book..571K, 2011Icar..214..477S}.

At high resolution, the spectral feature associated with orbital motion is revealed to be bimodal as a result of splitting between oscillations at different frequencies along different axes (Figure \ref{fig: zoomed Jovian field FFT at Europa}). The $B_{\mathrm{d}, z}$ peak is located at $84.61~\mathrm{hr}$, while $B_{\mathrm{d}, x}$ and $B_{\mathrm{d}, y}$ peak just below the $85.23~\mathrm{hr}$ orbital period at $85.20~\mathrm{hr}$.
These features are not reported in \citet{2011Icar..214..477S}, but can be understood by considering the different origins of $X$, $Y$, and $Z$ variations associated with orbital motion.

As mentioned above, Europa's orbital inclination ($i_\mathrm{Eur} = 0.47 \degree$) provides an additional source of oscillation in Europa's position relative to Jupiter's magnetic equator\textemdash in a Jupiter-centric frame rotating at Europa's orbital frequency, Jupiter's magnetic equator precesses at the synodic frequency, while Europa would appear to bob up and down, due to its inclination.
In the absence of precession, this vertical oscillation would occur at the orbital frequency, but nodal precession causes the points of maximum distance from the Jovian equatorial plane to drift, so that the oscillation occurs at a period of $(\omega_\mathrm{orb} - \omega_\mathrm{prec,nodal})^{-1} \approx 85.20 ~\mathrm{hr}$.
As with the synodic variation, this magnetic variation is primarily confined to the $X\text{-}Y$ plane of the IAU Europa system. 

Europa's radial distance to Jupiter also oscillates, owing to Europa's eccentric orbit ($e_\mathrm{Eur} = 0.009$). For a dipolar field, the variation in intensity from periapsis to apoapsis is $[(1-e_\mathrm{Eur})/(1+e_\mathrm{Eur})]^3 \approx 0.95$. However, as with the effect of inclination, this does not occur at precisely the orbital period, due to the effects of apsidal precession. 
Europa's periapsis precesses at an average rate of $\omega_\mathrm{prec, aps} = -0.7395 \degree ~\mathrm{day}^{-1}$, yielding an average time between periapses of $(\omega_\mathrm{orb} - \omega_\mathrm{prec, aps})^{-1} \approx 84.61~\mathrm{hr}$. The combined effects of apsidal and nodal precession explain the twin-peaked structure in the magnetic spectrum near the orbital period.

Similar peak splitting occurs at the beat frequencies of the synodic and orbital frequencies shown in Figure \ref{fig: Jovian field FFT at Europa}. In these cases the synodic oscillation is modulated by both Europa's radial distance from Jupiter and Europa's position with respect to the sub-solar Jovian longitude. These modulation frequencies are slightly faster and slower, respectively, than the orbital frequency, resulting in bimodal sidebands flanking the synodic oscillation and its harmonics (Figure \ref{fig: zoomed Jovian field FFT at Europa}). At high induction efficiencies ($A \sim 1$), each source of magnetic variation with a magnitude ${\gtrsim}1~\mathrm{nT}$ could plausibly produce an induction response detectable by a spacecraft magnetometer.

\subsection{Europa's Induced Field}
\label{sec: europa induction}
\begin{figure}
	\centering
	\includegraphics[width=0.5\textwidth]{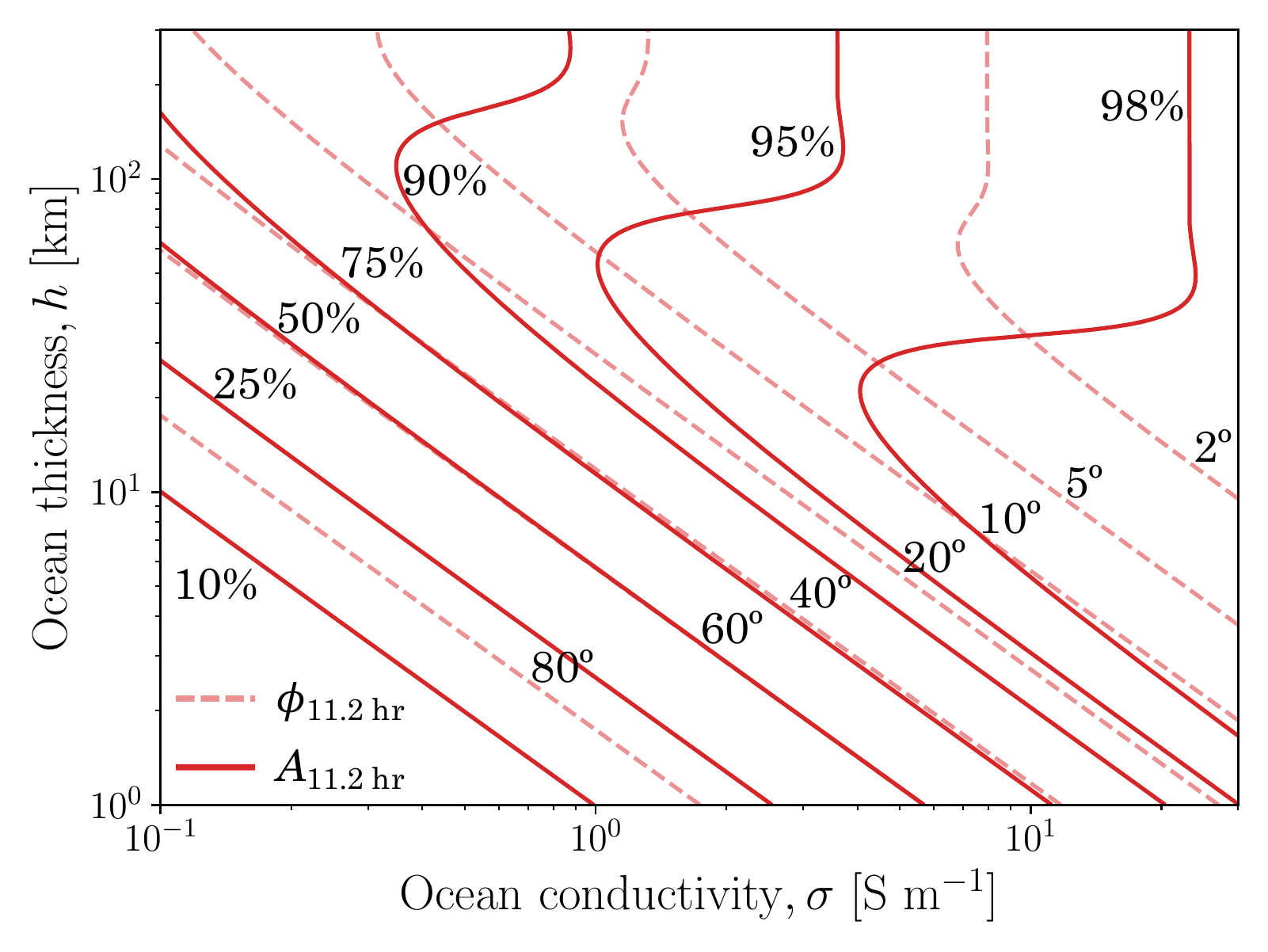}
	\caption{\small Contours of induction amplitude and phase lag in response to a driving magnetic field at the synodic period ($11.2~\mathrm{hr}$) as a function of ocean conductivity ($\sigma$) and thickness ($h$), after \citet{2000Icar..147..329Z}. Assumed ice shell thickness is $d=0~\mathrm{km}$. The amplitude of the synodic driving field is ${\sim}200~\mathrm{nT}$.}
	\label{fig: synodic contour}
\end{figure}

\begin{figure}
	\centering
	\includegraphics[width=0.5\textwidth]{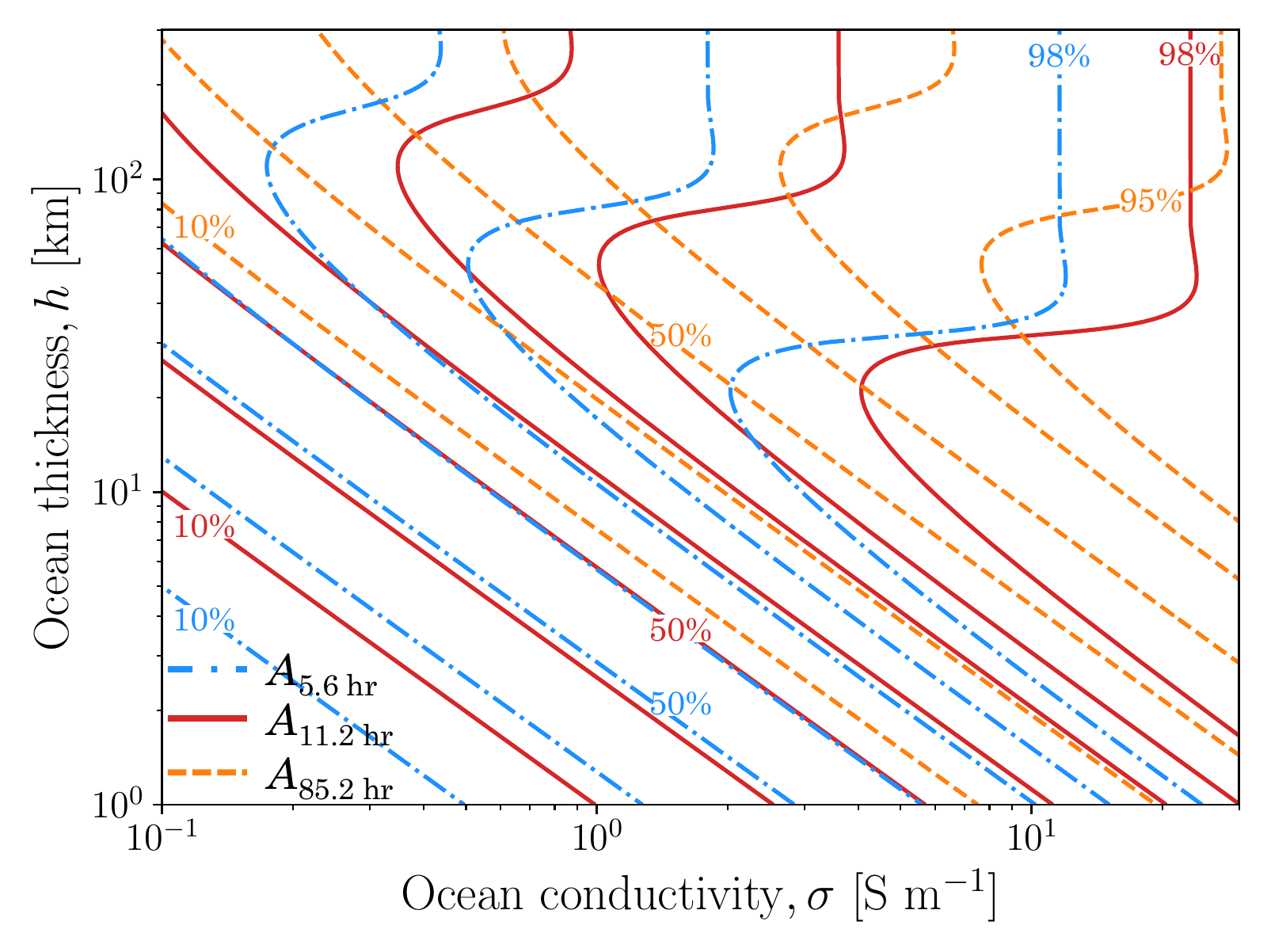}
	\caption{\small Contours of induction amplitude in response to a driving magnetic field at multiple frequencies as a function of ocean conductivity ($\sigma$) and thickness ($h$). The response at the synodic period ($11.2~\mathrm{hr}$) is given by the red solid line, at the half synodic period ($5.6~\mathrm{hr}$) by the blue dash-dotted line, and at the orbital period ($85.2~\mathrm{hr}$) by the orange dashed line. Contours are shown at amplitudes of $10, 25, 50, 75, 90, \text{ and } 95\%$ for all periods and at $98\%$ for the $5.6$ and $11.2~\mathrm{hr}$ oscillation. Assumed ice shell thickness is $d=0~\mathrm{km}$.}
	\label{fig: multifreq contour}
\end{figure}

The induced magnetic field measured by a spacecraft during a Europa flyby is determined by Europa's internal structure and the amplitudes and frequencies of the driving field. We calculated the induction amplitude and phase lag for a range of plausible interiors at the key driving frequencies identified above (Figures \ref{fig: synodic contour} and \ref{fig: multifreq contour}), obtaining results consistent with previous studies \citep[e.g.,][]{2000Icar..147..329Z, 2002AsBio...2...93K, 2021JGRE..12606418V}. The contour plots illustrate the degeneracy in inferring ocean structure from a single-frequency measurement of induction efficiency; a single value of $A$ can be produced by a family of plausible subsurface oceans. This explains the broad bounds on ocean parameters obtained from the Galileo data. \citet{2000Icar..147..329Z} found $70\% < A_{11.2~\mathrm{hr}} < 100\%$, corresponding to a large fraction of the plotted parameter space. Even the tightly constrained value of $A_{11.2~\mathrm{hr}} = 97 \pm 2 \%$ from \citet{2004JGRE..109.5006S} spans large ranges of plausible values of ocean thickness and conductivity.

In areas where contour lines from two different quantities cross, measuring both can, in principle, determine a unique solution for the Europan ocean. However, when contour lines do not have a unique point of intersection, the degeneracy is not broken. 
This can occur when the ocean is either much shallower or much deeper than the skin depth, $s = (\mathrm{\mu}_0 \sigma \omega / 2)^{-1/2}$, for all the frequencies being considered.
In situations with $h \ll s$ or $h \gg s$, contour lines are roughly parallel. Degenerate solutions can also occur when contours are not parallel but intersect at multiple points, as is the case with the $A = 98\%$ contour lines for the synodic and orbital periods (Figure \ref{fig: multifreq contour}). Expanding the range of measured frequencies so that $h \sim s$ for at least one frequency and leveraging both amplitude and phase information can resolve these uncertainties. By conducting more than 40 flybys of Europa, ECM will measure the induction response of Europa at wide range of periods, enabling ocean characterization over a large swath of plausible parameter space.

%%%%% Section: Simulating ECM
\section{Simulated Europa Clipper Magnetometer Data}
\label{sec: clipper magnet}
To evaluate the performance of ECM, we simulated magnetometry for each Europa flyby in a proposed tour (21F31v1) and considered a range of plausible interior structures. 
For each flyby we simulated data when Europa Clipper was within $10 \reuropa$ of Europa and inverted data sampled at $30~\mathrm{sec}$ cadence.
The model used to generate these data describes the Jovian magnetosphere, Europa's inductive response, and multiple error sources\textemdash principally systematic and random sensor noise and errors associated with incomplete removal of moon-plasma interaction fields.

The ambient magnetic field is determined using the Jovian magnetosphere model described in Section \ref{sec: jovian field} and the induction response from the three-layer internal structure model (Section \ref{sec: europa induction}). We used the full spectrum of magnetic variability at Europa to drive magnetic induction. This spectrum was calculated using a time series of the magnetic field at Europa that spans Europa Clipper's planned flybys of Europa (${\sim}3 ~\mathrm{yr}$) with $20~\mathrm{min}$ cadence, yielding a spectrum with ${\sim}40,000$ frequencies.

\begin{figure}
	\centering
	\includegraphics[width=0.5\textwidth]{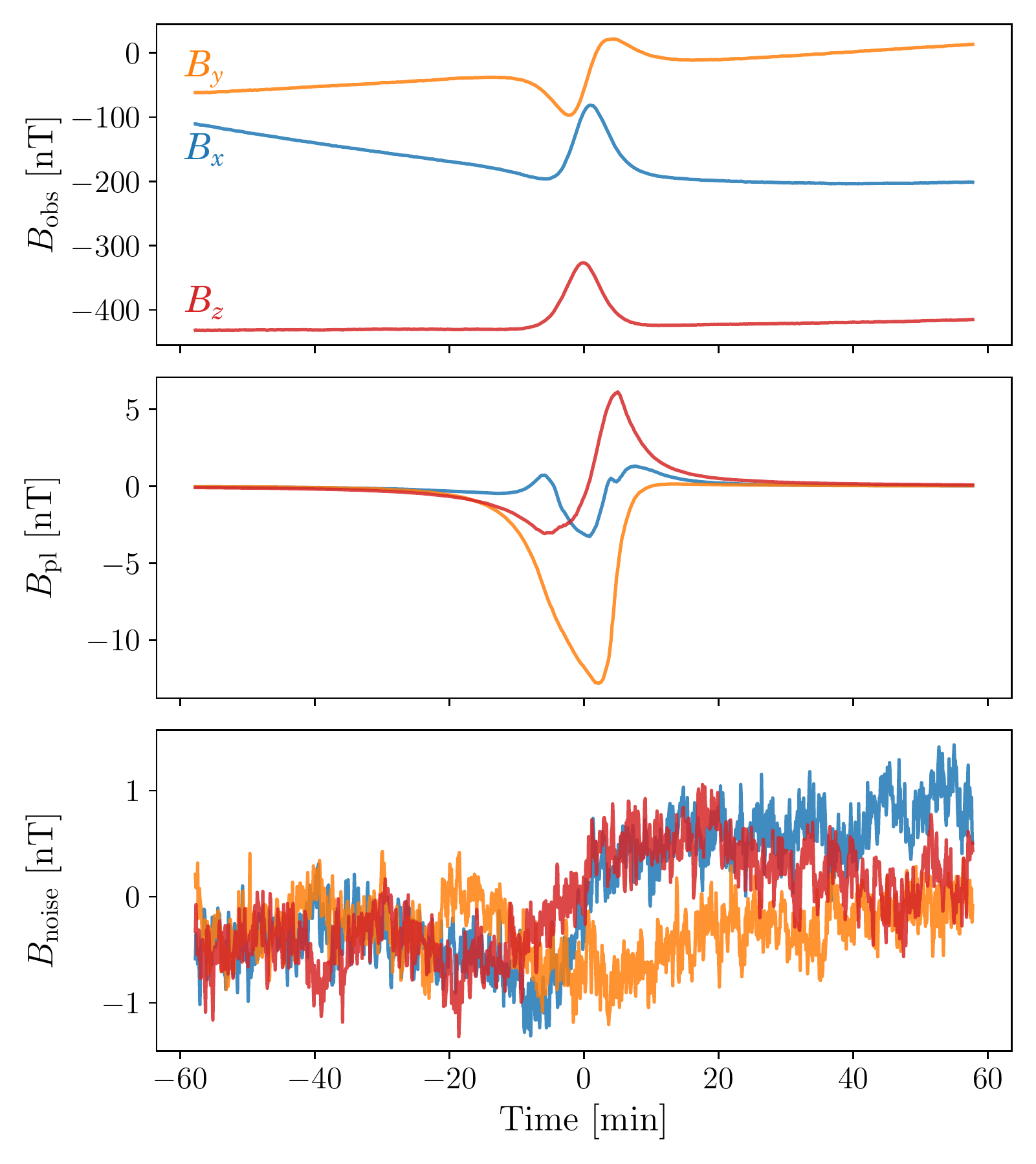}
	\caption{\small Components of simulated magnetic ECM data. Shown are simulated ECM data (top), moon-plasma interaction artifacts (middle), and sensor noise (bottom) during a flyby of Europa as a function of time relative to closest approach at five second cadence. The $B_x$, $B_y$, and $B_z$ components of each field are shown in blue, orange, and red, respectively. Plasma artifacts are at only $10\%$ of the original moon-plasma interaction field intensity, but remain a considerable source of systematic error.}
	\label{fig: simulated ecm data}
\end{figure}

\subsection{Moon-Plasma Interaction}
\label{sec: mpi}
Europa's interaction with the plasma in the Jovian magnetosphere perturbs both the local magnetic field and plasma environment \citep[e.g.,][]{2009euro.book..545K}. The ambient plasma conditions are time-variable, modulated by Jupiter's ${\sim}11.2~\mathrm{hr}$ synodic rotation, while the interaction with Europa depends on the state of Europa's tenuous atmosphere, which in turn varies with the ambient plasma conditions \citep[e.g.,][]{2020JGRA..12527485B}. The result of this non-linear interaction is a variable magnetic perturbation that is typically more pronounced when Europa is closest to the plasma sheet \citep[e.g.,][]{2020JA028888}. Importantly, these perturbations are comparable in scale to $\reuropa$ and can mask or mimic the signal from the induced dipole (Figure \ref{fig: simulated ecm data}, middle).

During Europa Clipper operations, it is expected that ECM and the Plasma Instrument for Magnetic Sounding (PIMS) will provide crucial measurements to initialize magnetohydrodynamic (MHD) simulations that can be used to quantitatively characterize the plasma interaction fields, allowing for accurate recovery of the induced field. We have simulated this process by injecting MHD-modeled plasma interaction fields, $\boldsymbol{B_\mathrm{pl}}$, into our mock data, scaling $\boldsymbol{B_\mathrm{pl}}$ by a factor of $0\text{--}1$ to represent incomplete removal of these perturbations. The simulated plasma fields are extracted from a catalog of multi-fluid MHD simulations built using the framework described in \citet{2020JA028888} that spans the likely conditions at Europa.

The catalog consists of 36 steady-state simulations covering the variation in Jupiter's magnetospheric magnetic field and plasma conditions over one ${\sim}11.2$-hour synodic period. Each simulation is initialized to represent the plasma interaction at evenly-spaced values of Europa's System III (S-III) longitude, a coordinate which rotates with Jupiter and is used here to determine Europa's position in Jupiter's magnetosphere. 
The MHD simulation uses the appropriate Jovian magnetic field values from the model of \citet{1997JGR...10211295K} and a model for the magnetospheric plasma density based on those presented by \citet{2015Icar..261....1B}. Europa's induced field is approximated assuming perfect induction efficiency ($A=100\%$) of the radial and azimuthal components of the magnetospheric field in each simulation. Three different models are used to represent Europa's atmosphere to account for the potential effects of production of neutral $\mathrm{O_2}$ due to sputtering of magnetospheric plasma against Europa's surface: one with low column density ($2.5 \times 10^{14}~\mathrm{cm^{-2}}$) corresponding to the situation where Europa is outside of Jupiter's central plasma sheet, another with high column density ($7.5 \times 10^{14}~\mathrm{cm^{-2}}$) for the case where Europa is near the center of Jupiter's plasma sheet, and another with intermediate column density ($5.0 \times 10^{14} ~\mathrm{cm^{-2}}$) for cases where Europa is transitioning towards or away from the plasma sheet.

Each Europa Clipper encounter occurs at a particular {S-III} longitude bracketed by two simulations, both of which are used to model the plasma magnetic field for the encounter. For example, if an encounter will occur when Europa is at $24 \degree$ {S-III} longitude, the catalog contains simulations at $20 \degree$ and at $30 \degree$. To approximate the plasma magnetic fields at points between the simulations, the simulated magnetic field was first extracted along the encounter trajectory from both neighboring simulations. The plasma perturbation fields were then calculated for each simulation by subtracting the induced and magnetospheric magnetic fields from the extracted total field. Then the plasma magnetic fields from the two simulations were weighted according to their proximity to the encounter in S-III longitude, and finally averaged to produce model plasma magnetic fields for that particular encounter. Even with $90\%$ of the plasma perturbation removed, ${\sim}10 ~\mathrm{nT}$-scale structures in the magnetic field remain in some flybys, potentially complicating the retrieval of the induced dipole.

\subsection{Noise and Contamination}
\label{sec: sim noise}
Our simulated magnetometry additionally included sensor noise and possible contamination from incomplete removal of sensor offsets and the spacecraft magnetic field.
Sensor noise along each axis of the magnetometer was modeled as a combination of flicker and white noise floor \citep[e.g.,][]{2016SSRv..199..189R}, defined by an amplitude spectral density of 
\begin{align}
100~\mathrm{pT} ~\mathrm{Hz}^{-1/2} \times \left( \frac{1 ~\mathrm{Hz}}{f} \right)^{-1/2} + 30~\mathrm{pT} ~\mathrm{Hz}^{-1/2} \text{,}
\end{align}
where $f$ is the frequency. We generated noise time series from white noise vectors that were transformed into the frequency domain, scaled by the the desired power law, and then returned to the time domain.

The measurement of the ambient field is systematically biased by calibration errors and the spacecraft's own magnetic field. In flight, these effects will be ameliorated through the use of gradiometry and spacecraft calibration rolls, which are expected to produce residual errors of ${\lesssim}1~\mathrm{nT}$ \citep[e.g.,][]{2004SSRv..114..331D, 2016AGUFMSM21A2455B}. We modeled the error after these corrections by adding a linearly drifting offset to the simulated data given by
\begin{align}
\boldsymbol{B_\mathrm{off}} = \boldsymbol{B_{\mathrm{off}, 0}} + \boldsymbol{B_\mathrm{drift}} (t - t_0) \text{,}
\end{align}
where the components of the initial offsets and drift rates were independently drawn from $\mathcal{U}{(-0.5, 0.5)}~\mathrm{nT}$ and $\mathcal{U}{(-1, 1)}~\mathrm{pT}~\mathrm{d}^{-1}$, respectively, and $t_0$ represents the time since the last calibration, conservatively assumed to be the beginning of the tour of the Jovian system. Unlike the other components of the simulated data, these noise sources are generated in a reference frame fixed with respect to the spacecraft, such that their signature in the simulated data set depends additionally on Europa Clipper's attitude during flybys. 
The resulting ${\sim}1 ~\mathrm{nT}$ structured noise  is $10\text{--}100$ times smaller than expected moon-plasma interaction residuals (Figure \ref{fig: simulated ecm data}, bottom), but still comparable to several percent of the amplitude of the driving field at the orbital and half synodic periods, potentially complicating accurate retrieval of the interior structure.

%%%%% Section: Bayesian retrieval
\section{Bayesian retrieval method}
\label{sec: bayesian retrieval}
We developed a technique for retrieving ocean structure\textemdash ice thickness, ocean thickness, and ocean conductivity\textemdash from spacecraft magnetometry using the framework of Bayesian inference. This method allows for self-consistent recovery of the ocean structure from the induction response at multiple frequencies and incorporates information from both the induction amplitude and phase delay. The resulting estimates of the ocean parameters include uncertainties from both measurement noise and the inherent degeneracy in the inversion problem.

\subsection{Retrieval Model}
\label{sec: retrieval model}
Parameter estimates are derived from fitting a retrieval model to spacecraft magnetometry data.
While some elements are shared, this model is distinct from the data generation model described in Section \ref{sec: clipper magnet}. In some cases, such as the number of magnetic oscillation frequencies considered, differences are due to computational limits. In others, however, we intentionally use lower-fidelity models to better mimic the process of recovering the interior from real ECM data.

Our retrieval model has three main components. The first is a model of the magnetic field in the vicinity of the moon. We assume the planetary field can be considered spatially uniform within a distance of several $\rmoon$ from the icy moon. The ambient field is then decomposed into a static background field ($\boldsymbol{B_0}$) and a series of sinusoidally time-varying fields at different frequencies,
\begin{align}
\label{eq: retrieval planet field}
\boldsymbol{B_\mathrm{amb}} = \boldsymbol{B_0} + \sum_{\omega, \oschat \in \left[ \unitvec{x}, \unitvec{y}, \unitvec{z} \right] }B_{\mathrm{d},\omega, \oschat} e^{-i ( \omega t + \phi_{\omega, \oschat}) } \oschat \text{,}
\end{align}
where $\phi_{\omega, \oschat}$ provides the phase of the driving oscillation at frequency $\omega$ in the $\oschat$ direction at a reference time. We allow only the frequency, $\omega$, to be fixed in Equation \eqref{eq: retrieval planet field}. The exact Jovian magnetospheric field driving the induction response will not be known during Clipper operations, so we treat the static field components as well as the amplitudes and phases of the driving field as free parameters. This results in $3 \times (2 n_\omega + 1)$ free parameters, where $n_\omega$ is the number of frequencies selected for the inversion and the $+1$ term comes from the static field.
\begin{deluxetable}{cl}[t]
\tablecaption{Total Field Model Parameters}
\label{table: parameters}
\tablehead{
\colhead{Parameter} & \colhead{Description}}
\startdata
$\boldsymbol{B_0}$ & Static background field at Europa (vector) \\
$B_{\mathrm{d},\omega, \oschat}$ & Amplitude of oscillation along $\oschat$ at frequency $\omega$ \\
$\phi_{\omega, \oschat}$ & Phase of oscillation along $\oschat$ at frequency $\omega$ \\
$d + h$ & Total hydrosphere thickness (ice + ocean) \\
$d$ & Ice shell thickness \\
$\sigma$ & Ocean conductivity \\
$\boldsymbol{B_{\mathrm{off}, 0}}$ & Initial sensor offsets (vector) \\
$\boldsymbol{B_\mathrm{drift}}$ & Linear drift rate (vector) \\
$\sigma_\mathrm{noise}$ & Magnetometer jitter \\
\enddata
\tablecomments{The total number of parameters is $6 n_\omega + 13$, where $n_\omega$ is the number of frequencies used in the inversion.}
\end{deluxetable}

\begin{figure}
	\centering
	\includegraphics[width=0.5\textwidth]{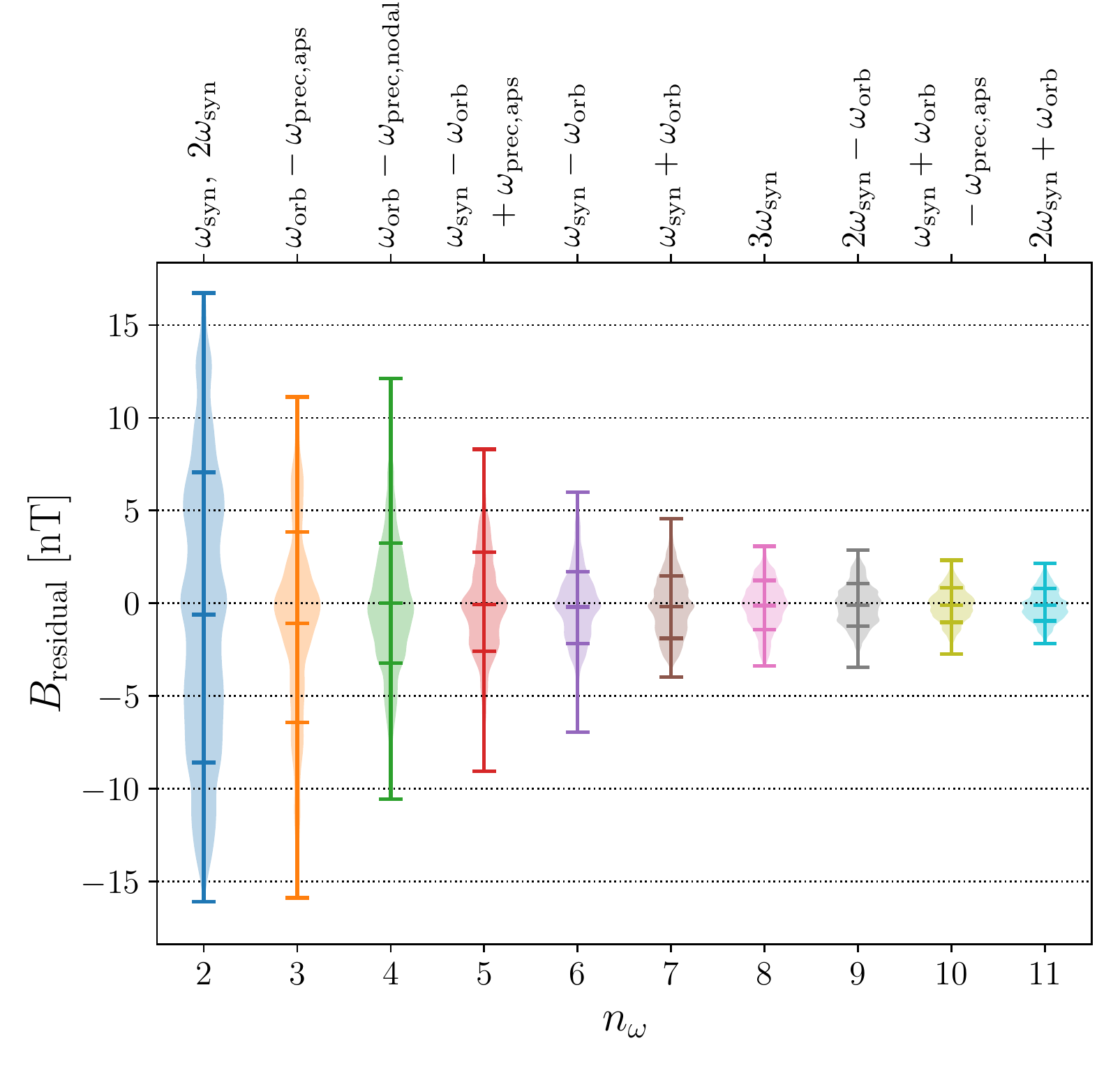}
	\caption{\small Accuracy of Jovian field model used in retrievals. Shown is a violin plot illustrating the accuracy of models with different numbers of modeled frequencies, $n_\omega$, compared with the high-fidelity model used to generate synthetic data (Section \ref{sec: jovian field}). Labels along the top axis indicate which frequencies were added to the model. Model residuals, $B_\mathrm{residual}$, are calculated for the predicted magnetic field at Europa during Clipper flybys. Shaded regions represent the smoothed distribution of model residuals, while horizontal lines indicate the minimum, maximum, mean, and the 16th and 84th percentiles of the residuals. Adding additional frequencies reduces systematic errors, and yields smaller means and variances for the residual distribution.}
	\label{fig: jupiter model accuracy}
\end{figure}

We tested the accuracy of representing the Jovian field with a small number of frequencies by calculating the field at Europa during Clipper flybys using the high-fidelity Jovian field model (Section \ref{sec: jovian field}) and comparing the results to the field obtained from the sparse frequency model in Equation \eqref{eq: retrieval planet field}. We define $B_\mathrm{residual}$ as the difference, by component, between the two models. The terms in the frequency series were determined from a simulated spectrum of the Jovian field at Europa (e.g., Figure \ref{fig: Jovian field FFT at Europa}). We progressively added frequencies in decreasing order of their total oscillation amplitude, beginning with the synodic and orbital frequencies and ending with all 11 frequencies with amplitudes ${\gtrsim}1 ~\mathrm{nT}$. As the number of frequencies increases, the accuracy of the retrieval model field improves, with diminishing improvements when more than eight frequencies are included (Figure \ref{fig: jupiter model accuracy}). We used nine frequencies to model the driving field in most of our analysis, as a balance between computational complexity and retrieval accuracy. For nine modeled frequencies, the mean and standard deviation of the residuals are reduced to levels comparable to other expected noise sources (Section \ref{sec: sim noise}), $\mu_\mathrm{res} \approx 0.1 ~\mathrm{nT}$ and $\sigma_\mathrm{res} \approx 1.1 ~\mathrm{nT}$, respectively.

The assumption of spatial uniformity also limits the accuracy of the retrieval model. For example, when Europa Clipper is $5 R_\mathrm{Eur}$ from Europa in the radial direction ($+X$ in IAU Europa) the resulting error is ${\sim}20 ~\mathrm{nT}$, comparable to the magnitude of oscillation at the orbital frequency. To account for this, we use a model of the planetary magnetic field ($B_\mathrm{pmf}$) to derive a correction term which represents the difference between the model field at the spacecraft and the moon: $\boldsymbol{\delta{B}} = \boldsymbol{B_\mathrm{pmf}}{(\boldsymbol{r_\mathrm{sc}})} - \boldsymbol{B_\mathrm{pmf}}{(\boldsymbol{r_\mathrm{Eur}})}$. 
To better simulate the process of inverting ECM data, where any magnetosphere model will only approximate the observed field, we use a simple dipole field model \citep[from][]{2018GeoRL..45.2590C} to calculate the correction term rather than using the high fidelity model in Section \ref{sec: jovian field}. We find that this typically accounts for ${\sim}90\%$ of the error introduced by the assumption of a spatially uniform field near Europa.

After the Jovian field, the second component of the retrieval model is the induced field. Adopting the three-layer internal structure model, an induced dipolar field is generated for each frequency and axis of the driving field according to Equation \eqref{eq: induced field}. Once the components of the driving field are specified (Equation [\ref{eq: retrieval planet field}]), this depends only on the three ocean properties ($d$, $h$, and $\sigma$). Because static gravity measurements can constrain the total water layer thickness, $d + h$, \citep[e.g.,][]{1998Sci...281.2019A}, we re-parameterize the internal structure model to make the total water layer thickness a free parameter. The resulting free parameters describing the ocean structure are then ($d, d+h, \sigma$).

The final component of the retrieval model is magnetic contamination from spacecraft fields and sensor noise. 
This is represented by both white noise and a model of drifting sensor offsets.
The white noise is assumed to have zero mean and is parameterized by a standard deviation $\sigma_\mathrm{noise}$, assumed to be equal in each axis of the fluxgate magnetometer. The offsets in each direction are independent and assumed to drift linearly over the course of the mission, $\boldsymbol{B_\mathrm{off}} = \boldsymbol{B_{\mathrm{off}, 0}} + \boldsymbol{B_\mathrm{drift}} t$. Unlike the previous components, these are computed in the spacecraft coordinate system and then rotated into the moon's body-fixed coordinate system using spacecraft attitude information encoded in SPICE kernels provided by the Europa Clipper project. The total field given by the retrieval model is then 
\begin{multline}
\label{eq: retrieval field}
\boldsymbol{B} = \boldsymbol{B_0} + \sum_{\omega, \oschat = \unitvec{x}, \unitvec{y}, \unitvec{z}} \left(B_{\mathrm{d},\omega, \oschat} e^{-i ( \omega t + \phi_{\omega, \oschat}) } \oschat  + \boldsymbol{B_{\mathrm{i}, \omega, \unitvec{e}}} \right)
\\
+ \boldsymbol{B_\mathrm{off}} + \boldsymbol{\delta{B}} + \mathcal{N}{(0, \sigma_\mathrm{noise})} \text{,}
\end{multline}
where the induced field term, $\boldsymbol{B_{\mathrm{i}, \omega, \unitvec{e}}}$, is given by Equation \eqref{eq: induced field}.

The total field model is determined by $3 \times (2 n_\omega + 1) + 10$ free parameters, or $67$ parameters for $n_\omega = 9$ as in our baseline analysis (Table \ref{table: parameters}). Comparison between this model field and the observed magnetometry forms the basis for parameter retrieval.

\subsection{Parameter Retrieval}
\label{sec: parameter retrieval}
The retrieval model is fit to magnetometry data using the Markov Chain Monte Carlo (MCMC) sampler provided by \texttt{EMCEE} \citep{2013PASP..125..306F} with a combination of the affine-invariant \citep{2010CAMCS...5...65G} and differential evolution \citep{terBraak2008SaC} ensemble sampling strategies. 
The result is an estimate of the posterior probability distribution for all the parameters in the retrieval model given the observed magnetometer data. For a set of model parameters $\theta$ and magnetometer data $D$, the posterior is written $p(\theta | D) \propto p(D | \theta) p(\theta)$, where the first term on the right is the likelihood, the second is the prior probability, and the proportionality is given by Bayes' theorem. The likelihood depends on how well the model fits the observed data. Under our assumption of independent, normally distributed noise with standard deviation $\sigma_\mathrm{noise}$, the likelihood is ${\propto}\exp{(-\chi^2 / 2)}$, where $\chi^2$ is the chi-squared statistic.

In most of our retrievals we adopt relatively uninformative priors. For the amplitudes and phases of the driving field we adopt uniform priors, centered around an estimate obtained from a model-generated time series. The width of the amplitude is the larger of $20\%$ of the estimated amplitude and $20~\mathrm{nT}$, and the width of the uniform phase prior is $\pi/2$. The initial sensor offset and drift rates are likewise uniform, $\boldsymbol{B_\mathrm{off}} \sim \mathcal{U}{(-1 , 1)}~\mathrm{nT}$ and $\boldsymbol{B_\mathrm{drift}} \sim \mathcal{U}{(-1 , 1)}~\mathrm{pT} ~\mathrm{day}^{-1}$. 

For the ocean parameters, we adopt broad priors that are consistent with existing constraints. \citet{2005Icar..177..397B} review estimates of the ice shell thickness derived from a variety of geological evidence, finding literature values of ${<}1$ to ${>}30~\mathrm{km}$, while \citet{2007Icar..189..424H} suggest that the high measured induction efficiency \citep{2004JGRE..109.5006S} requires a thin ice shell, ${<}15~\mathrm{km}$. 
We therefore take $d \sim \mathcal{U}{(0, 50)}~\mathrm{km}$ as our prior on the ice shell thickness. Gravity data from Galileo constrain the total water layer thickness to $80\text{--}170~\mathrm{km}$ \citep{1998Sci...281.2019A}, so we adopt a uniform prior over this range for $d + h$. 
Based on analysis \citep{2007Icar..189..424H} of the salinity ranges for $\mathrm{NaCl}$ and $\mathrm{MgSO_4}$ that are both physically possible and consistent with the induced field magnitude \citep[from][]{2004JGRE..109.5006S, 2000Icar..147..329Z}, we take $\log{[ \sigma / (1~\cond) ]} \sim \mathcal{U}{\left( \log{(0.07)},  \log{(30)} \right)}$ as our prior on the conductivity. These priors are intentionally wide and minimally informative so as to provide more conservative estimates of ECM's ability to recover Europa's interior. We examine the effect of more informative priors, derived from notional Europa Clipper static gravity measurements, in Section \ref{sec: informative priors}.

\subsection{Comparison with Previous Work}
Our approach differs from previous efforts to infer the internal structure of Europa using induction in several respects. Past analysis focused on recovering the induction amplitude at the synodic frequency from the Galileo data, making the internal structure necessarily degenerate \citep[][Section \ref{sec: europa induction}]{2004JGRE..109.5006S,2007Icar..192...41S,2000Icar..147..329Z}. Because Europa Clipper's mission will support multi-frequency induction measurements, our approach self-consistently incorporates the induction amplitude and phase at multiple frequencies. This allows us to directly obtain the ocean parameters from the magnetometry, rather than fitting only for the induction amplitude or other proxies of these parameters. Further, because we recover posterior distributions and not a single best-fit as in the case of least-squares fitting \citep[e.g.,][]{2004JGRE..109.5006S}, our estimates of the ocean parameters are robust, even in cases where the solution remains degenerate.

%%%%% Section: Results
\section{Results}
\label{sec: results}
We evaluated the ability of ECM to characterize the internal structure of Europa from simulated magnetometry using our Bayesian retrieval method for a range of representative scenarios. In each case, we simulated and analyzed an entire mission's worth of simulated data to recover the parameters of the retrieval model. We then compared the recovered parameters to the input parameters to assess ECM's performance and our retrieval method.
\begin{deluxetable}{cccc}[t]
\tablecaption{Europan Ocean Scenarios}
\label{table: scenarios}
\tablehead{
\colhead{Scenario} & \colhead{Ice, $d~\mathrm{[km]}$} & \colhead{Ocean, $h ~\mathrm{[km]}$} & Conductivity, $\sigma ~\mathrm{[\cond]}$}
\startdata
A & 10 & 120 & 0.1 \\
B & 2 & 163 & 2.75 \\ %B2
C & 30 & 50 & 10 \\
D & 20 & 80 & 27.5 \\
E & 30 & 80 & 0.1 \\
F & 2 & 163 & 27.5 \\ %F2
G & 15 & 100 & 1 % H
\enddata
\tablecomments{For each scenario we consider both a case with perfect and incomplete (90\%) removal of moon-plasma interaction fields.}
\end{deluxetable}
\begin{figure}
	\centering
	\includegraphics[width=0.5\textwidth]{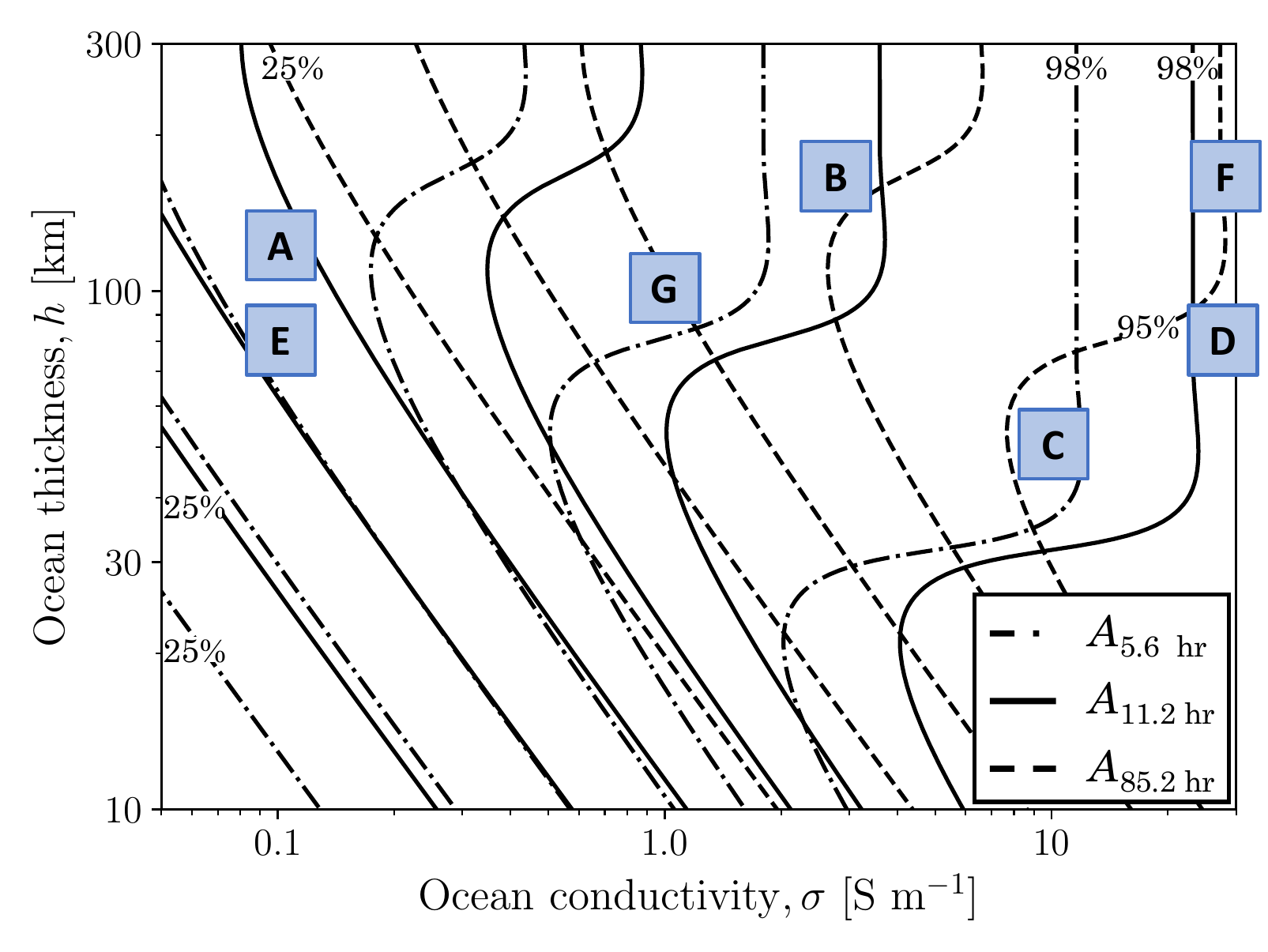}
	\caption{\small Europan interiors considered in our analysis. For reference, the induction response amplitudes at the three highest-amplitude oscillation frequencies are shown as a function of ocean conductivity and thickness, assuming no ice shell ($d = 0~\mathrm{km}$). Contours are shown at amplitudes of $25, 50, 75, 90, \text{ and } 95\%$ for all periods and at $98\%$ for the $5.6$ and $11.2~\mathrm{hr}$ periods. See Table \ref{table: scenarios}.}
	\label{fig: scenarios}
\end{figure}

We considered seven plausible ocean structures and two levels of moon-plasma interaction fields for each of these scenarios. 
The ocean structures were drawn from the parameter space described in Section \ref{sec: parameter retrieval}. Four samples were drawn using a Latin Hypercube approach \citep[e.g.,][]{10.2307/1268522}, to which we added three scenarios to represent the minimum and maximum induction response and one intermediate case (Table \ref{table: scenarios}). These scenarios span the range of plausible structures allowed by the Galileo data (Figure \ref{fig: scenarios}). For each of these internal structure models, we then simulated ECM data assuming complete or partial correction of moon-plasma interaction effects, adding moon-plasma interaction fields with intensities of $0$ and $10\%$ of their MHD model values.
Finally, the instrument noise (Section \ref{sec: sim noise}) used in each simulation and recovery is unique, so that the different scenarios also span a range of plausible noise levels.

\begin{figure}
	\centering
	\includegraphics[width=0.5\textwidth]{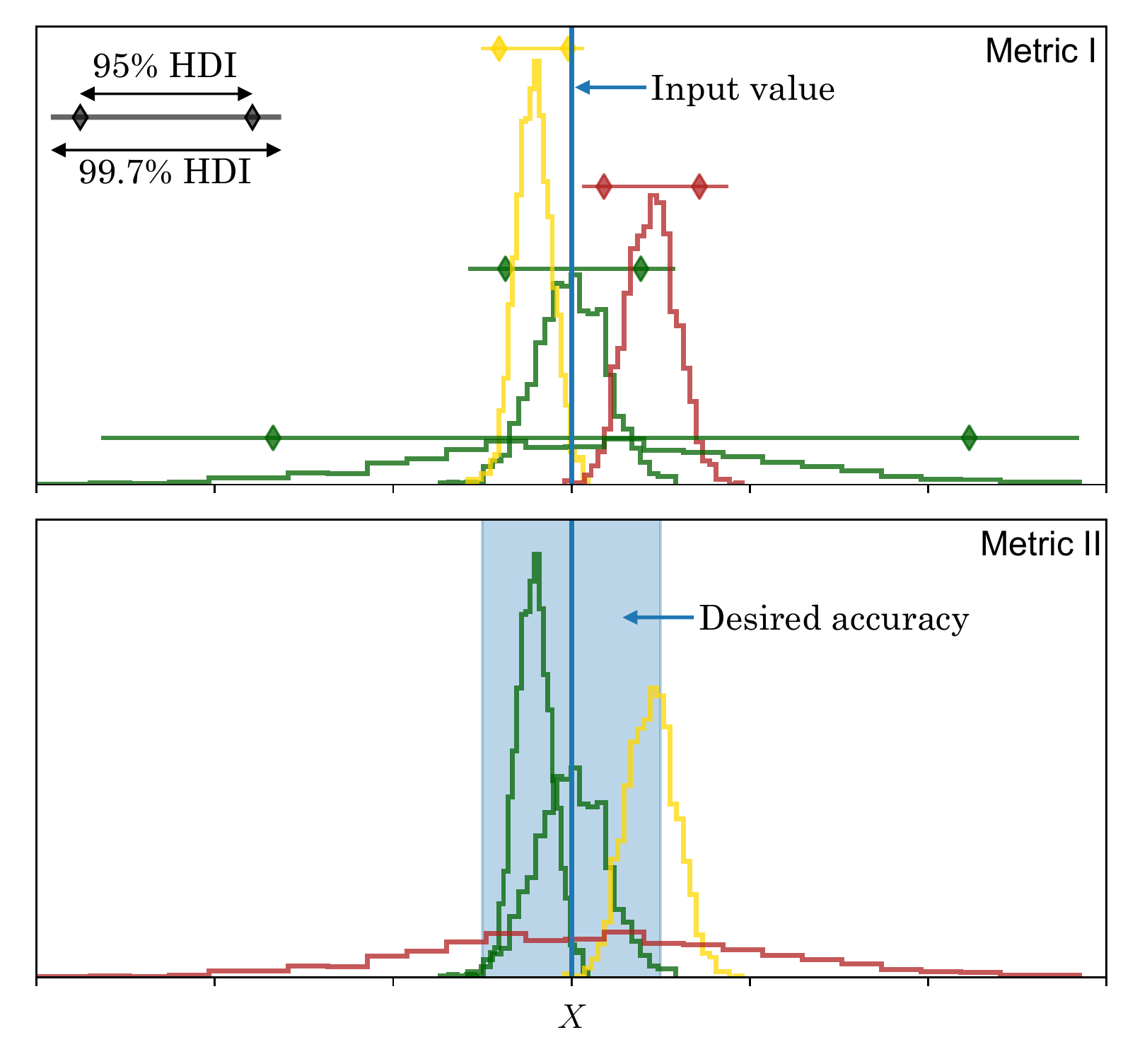}
	\caption{\small Illustration of metrics used for evaluation of simulated ECM data sets. Green, yellow, and red posterior probability distributions indicate successful, marginal, and failed recoveries of the notional parameter $X$, respectively. The distributions in both panels are identical, but only one fully passes both tests. \textbf{(Top)} Comparison of posterior highest-density interval (HDI) with true value. Successful, marginal, and failed parameter recoveries correspond to truth values lying inside the ${\le}95$, ${\le}99.7$, and ${>}99.7\%$ HDIs, respectively. The $99.7\%$ interval is indicated by the horizontal line over the distribution, while the diamonds indicate the bounds of the $95\%$ interval. \textbf{(Bottom)} Comparison of retrieved posterior with desired measurement accuracy. Probability masses of ${>}68$, ${>}50$, and ${\le}50\%$ define success, marginal, and failed retrievals.}
	\label{fig: metrics}
\end{figure}

We assessed recovery of the three ocean parameters in each scenario using two metrics. Metric I assesses accuracy by quantifying how well the retrieved posterior for an ocean parameter encompasses the input value. We calculate the 95th and 99.7th percentile highest-density intervals (HDI), the minimum-width Bayesian credible intervals containing the specified probability mass. When the true value of the ocean parameter falls within the 95\% interval, we designate the retrieval a success. Cases where the input value falls only within the 99.7\% interval are marginal, while those cases where the input lies outside the 99.7\% range are failures (Figure \ref{fig: metrics}). Importantly, this metric does not necessarily indicate that the ocean has been successfully characterized\textemdash our relatively uninformative priors satisfy it. Instead, it measures whether the uncertainty in the recovered parameters inferred from the posterior is appropriate. Failure may result from systematic bias in the posterior distribution, which itself may stem from significant un-modeled effects.

Metric II evaluates how well the ocean has been characterized. In this test, we calculate the total probability mass of the posterior contained within a specified interval around the input value. 
We require ${>}68\%$ of the probability mass to be within this interval for a successful retrieval, or ${>}50\%$ of the probability mass for a marginal recovery (Figure \ref{fig: metrics}). These thresholds are selected so that a marginal score indicates that the accuracy range includes the median of the posterior, while a success requires probability mass at least equivalent to the ${\pm}1\text{-}\sigma$ range of a Gaussian distribution to fall within the accuracy window. For the ice thickness and ocean conductivity, the accuracy range is given by $\pm 50\%$ of the input value to match the Europa Clipper science objectives \citep{9172447}. For the ocean thickness this would cover most of the allowed range for most of the modeled scenarios, so we adopt a more stringent requirement of $\pm 25~\mathrm{km}$ for this parameter.
Since our uninformative priors have only a small fraction of their probability mass near the correct values, they fail this metric. Conversely, recovered distributions with credible intervals excluding the true value may still pass if their probability mass is sufficiently concentrated near the input value. In this case, the inferred value is near the truth, but the parameter's uncertainty is underestimated.

These metrics could, in principle, be applied to all of the recovered parameters in the retrieval model (Table \ref{table: parameters}). However, because the goal of the ECM investigation is ocean characterization, in the following analysis we evaluate scenario performance based solely on the scores of the three ocean parameters.

\subsection{Baseline Performance}
\label{sec: baseline results}
\begin{figure}
	\centering
	\includegraphics[width=0.5\textwidth]{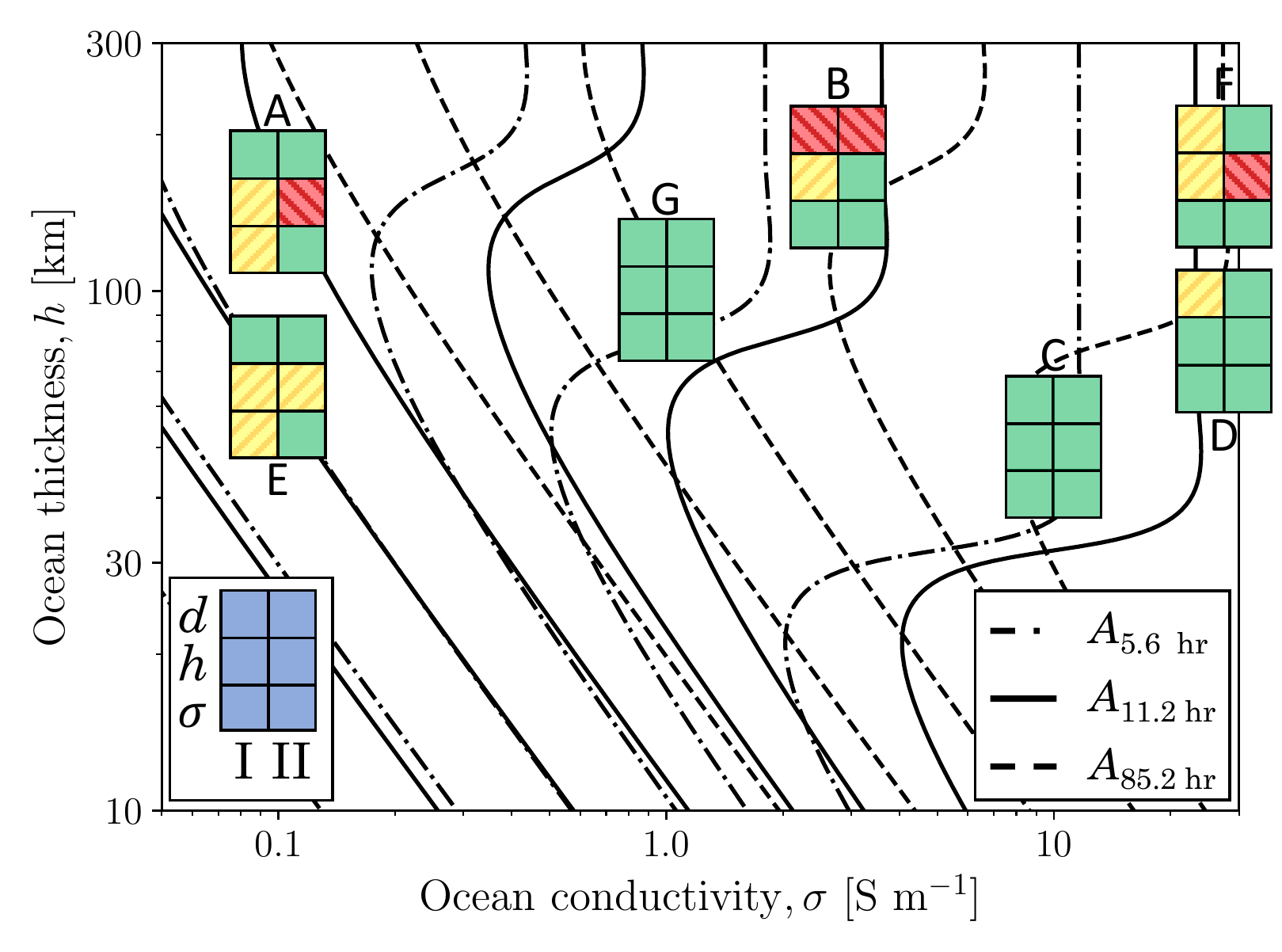}
	\caption{\small Baseline performance of Bayesian retrieval of Europan interior for all scenarios. Solid green boxes correspond to successful recovery of a parameter as measured by the indicated metric. Yellow boxes with positively sloped stripes indicate a marginal recovery and red boxes with negatively sloped stripes indicate a failure. Contours of constant induction response assuming no ice shell ($d=0$) are shown at amplitudes of $25, 50, 75, 90, \text{ and } 95\%$ for the three indicated frequencies and at $98\%$ for the $5.6$ and $11.2~\mathrm{hr}$ periods.}
	\label{fig: baseline results}
	% 29 success, 9 marginal, 4 failure ; 9 frequencies
\end{figure}

As a baseline, we ran retrievals on synthetic ECM data sets assuming complete removal of moon-plasma interaction effects. In this case, the effectiveness of parameter recovery is determined by the noise characteristics of the magnetometer, the assumed internal structure of Europa, and the timing and geometry of the spacecraft's Europa flybys. 
Across all scenarios, the three ocean parameters are recovered successfully or marginally for both metrics approximately 85\% of the time (Figure \ref{fig: baseline results}). The ocean conductivity is recovered successfully or marginally in all scenarios and the ice shell thickness is recovered in all but one (Scenario B). The ocean thickness poses more difficulty, generating failures in two scenarios (A and F).
\begin{figure*}
	\centering
	\includegraphics[width=\textwidth]{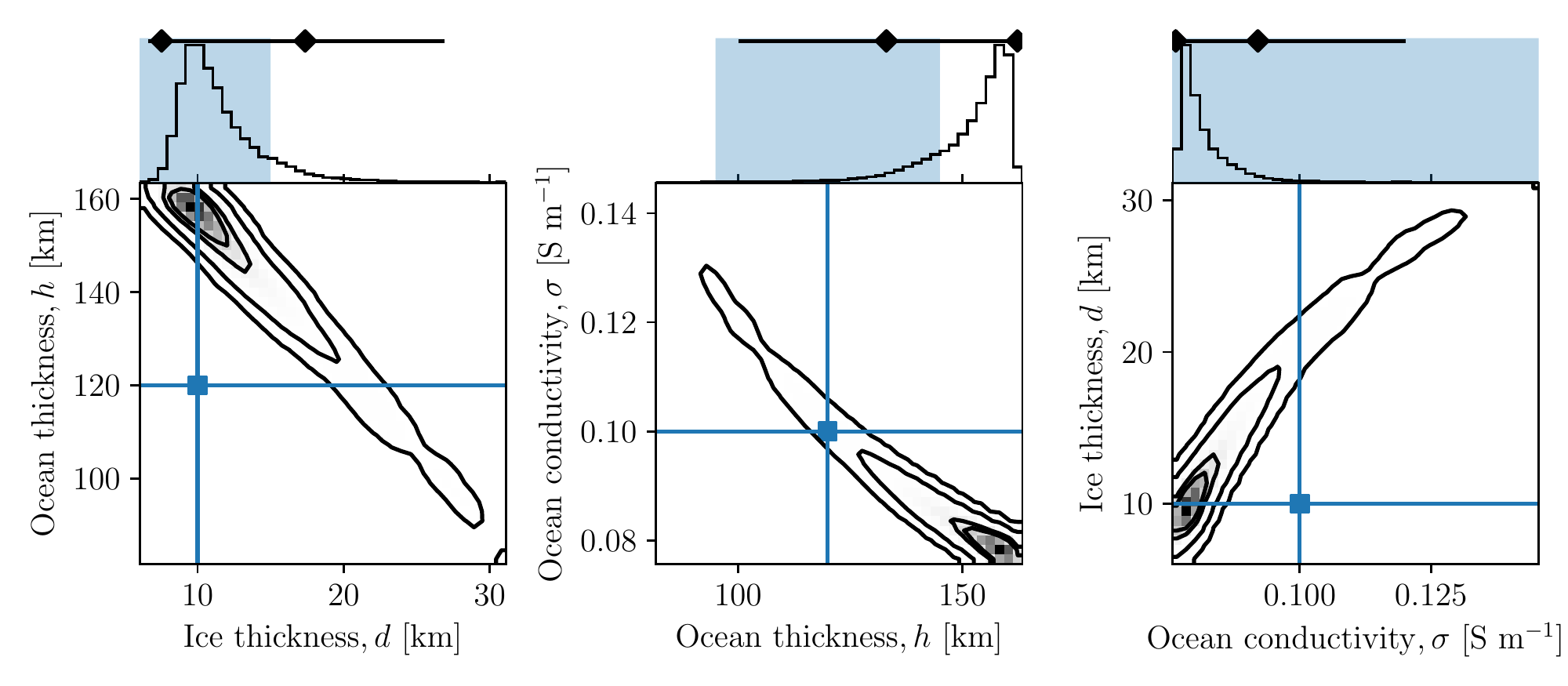}
	\caption{\small Example retrieval of Europan interior for Scenario A with 1D and 2D posterior distributions for the three ocean parameters and parameter pairings. Two-dimensional posteriors are illustrated by contours of constant probability density drawn enclosing $5, 50, 68, 95, \text{ and } 99.7\%$ of the probability mass. Blue lines and square markers indicate the input values. Above each contour plot, a histogram displays the posterior for the ocean parameter on the $x$-axis. In these plots, the horizontal black line indicates the $99.7\%$ interval, the black diamonds indicate the bounds of the $95\%$ interval, and the blue shaded region indicates the Metric II interval.}
	\label{fig: baseline scenario a}
\end{figure*}

Scenario A provides an example of a retrieval with successful or marginal ice shell and ocean conductivity recovery and a failed ocean thickness recovery (Figure \ref{fig: baseline scenario a}). The degeneracy between the three ocean parameters is not fully broken and for all three the uncertainty around the median or modal value is considerably asymmetric. While the ocean conductivity is tightly and accurately confined to the low conductivity region, the ice thickness and ocean thickness distributions feature long tails so that the range of plausible values at the 99.7\% level is large. For all three parameters the input value falls inside the 99.7\% interval or better, resulting in the marginal and successful scores on Metric I. 
The $\pm50\%$ accuracy windows for the ice shell and ocean conductivity capture large fractions of the total probability mass. Accordingly, the ocean conductivity and ice thickness are successfully recovered according to Metric II. By contrast, for the ocean thickness more than half the probability mass lies outside the accuracy window of $120 \pm 25~\mathrm{km}$, resulting in a failure for Metric II. 

\begin{figure*}
	\centering
	\includegraphics[width=\textwidth]{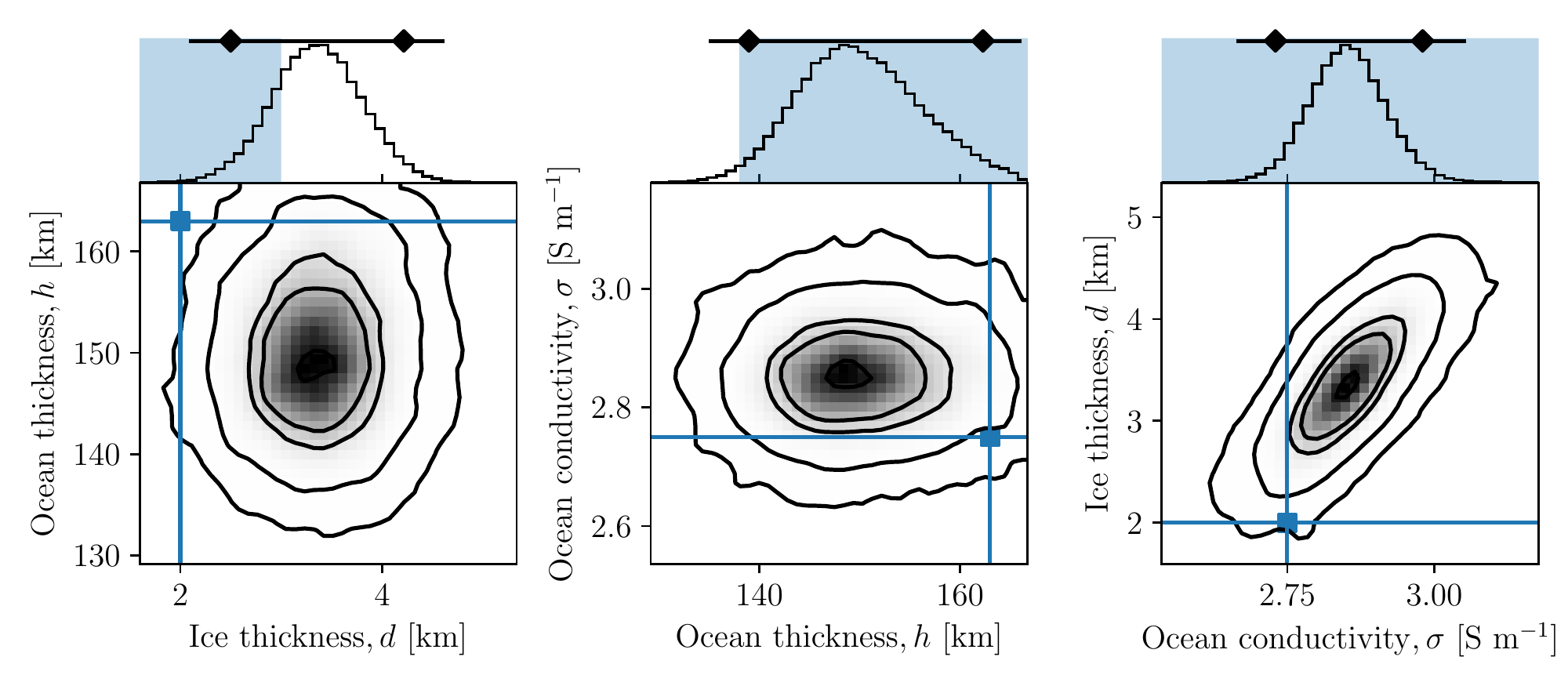}
	\caption{\small Example retrieval of Europan interior for Scenario B with 1D and 2D posterior distributions for the three ocean parameters and parameter pairings. Two-dimensional posteriors are illustrated by contours of constant probability density drawn enclosing $5, 50, 68, 95, \text{ and } 99.7\%$ of the probability mass. Blue lines and square markers indicate the input values. Above each contour plot, a histogram displays the posterior for the ocean parameter on the $x$-axis. In these plots, the horizontal black line indicates the $99.7\%$ interval, the black diamonds indicate the bounds of the $95\%$ interval, and the blue shaded region indicates the Metric II interval.}
	\label{fig: baseline scenario b}
\end{figure*}

In Scenario B, the ocean parameter degeneracies are largely broken and the resulting posteriors are narrow and more symmetric (Figure \ref{fig: baseline scenario b}). In particular, the ice thickness and ocean conductivity distributions appear nearly Gaussian, while the ocean thickness exhibits some asymmetry and has fatter tails. The ocean conductivity and thickness are successfully or marginally recovered according to both Metric I and II.
However, the ice shell recovery is a failure: the input value of $2~\mathrm{km}$ falls just outside of the 99.7\% credible interval, which spans $2.1\text{--}4.6~\mathrm{km}$ and the narrow accuracy window ($2 \pm 1~\mathrm{km}$) contains only ${\sim}20\%$ of the posterior probability mass. The relatively small absolute bias of the posterior distribution therefore results in failures on both ice shell metrics.

These two scenarios exhibit two different classes of failures, one related to correctable retrieval model deficiencies and the other dependent on the chosen scenario. The first type is illustrated by the failure to recover the ice shell thickness in Scenario B. In this case, the modestly biased posterior is the result of the retrieval model fitting to systematic noise in the magnetic field data. In particular, using a limited number of frequencies to model the driving field introduces errors in the recovered driving field parameters, which in turn produce errors in the recovered ocean parameters. At nine frequencies, artifacts persist at the ${\sim}1~\mathrm{nT}$ level (Figure \ref{fig: jupiter model accuracy}). By comparison, for a perfectly conductive ocean the normalized induction response is $A = (\rocean / \rmoon)^3 \approx (1 - 3 d / \rmoon)$. A difference of one kilometer in the ice shell thickness consequently changes the induction efficiency by ${\sim}0.2\%$, corresponding to ${\sim}0.5~\mathrm{nT}$ error at the synodic frequency. Bias in the recovered ice shell thickness at the ${\sim}1~\mathrm{km}$ scale is therefore likely to be a persistent feature, making our metrics challenging for thin ice shells, though this may be ameliorated as more frequencies are considered. Finally, obtaining reliable posteriors does not generally require errors ${<}1 ~\mathrm{nT}$. Rather, the bias exhibited in Scenario B occurs because the errors in the data were dominated by systematics at frequencies of significant magnetic oscillation. When these systematics do not dominate, accurate retrievals are possible, even in the presence of increased noise (see Section \ref{sec: mpi retrievals}).

\begin{figure}
	\centering
	\includegraphics[width=0.5\textwidth]{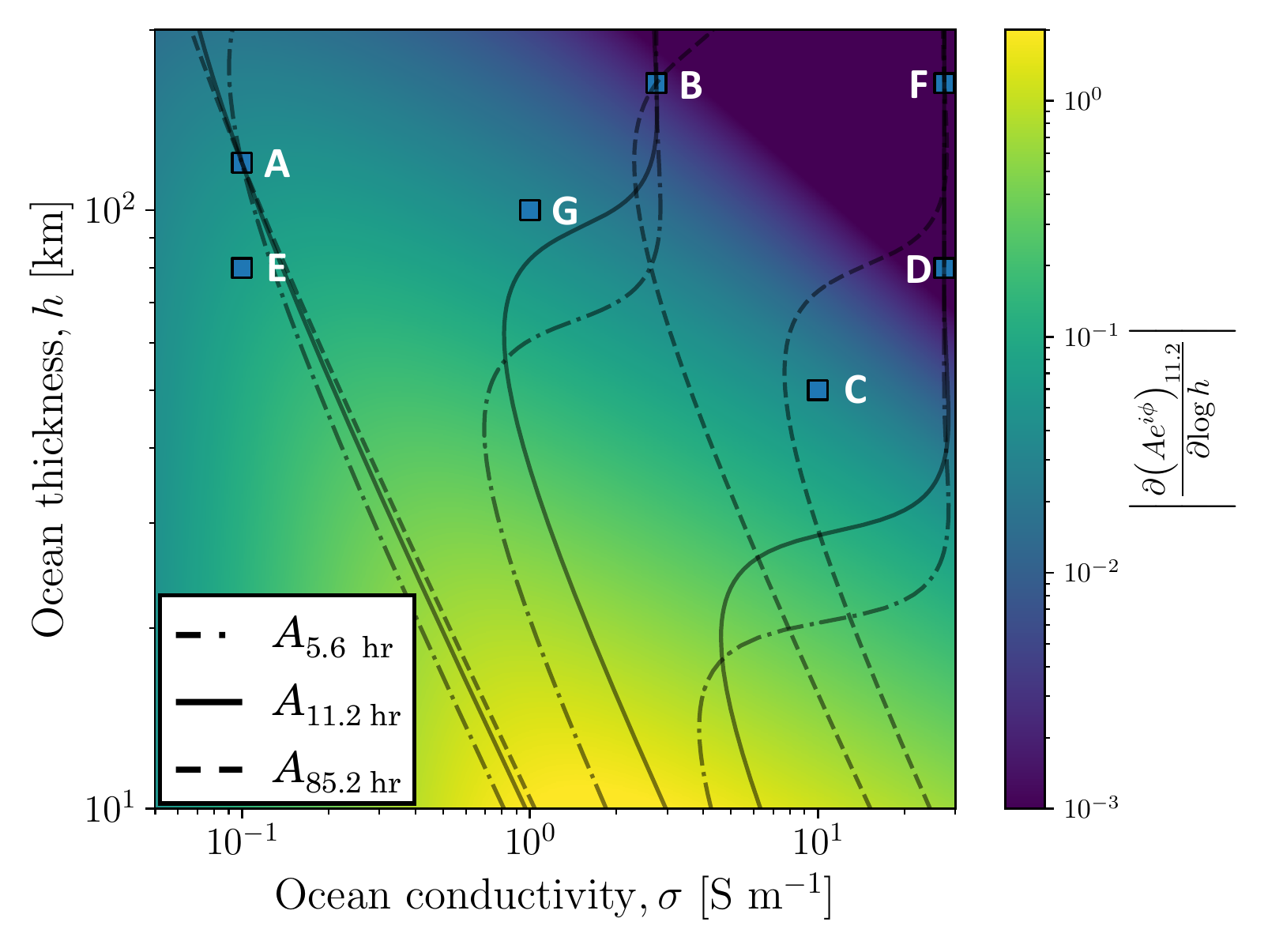}
	\caption{\small Sensitivity of the induction response to ocean parameters. Shown is the derivative of complex induction amplitude ($A e^{i \phi}$) at the synodic frequency with respect to fractional change in ocean thickness as a function of ocean conductivity and thickness. Scenarios are indicated by blue boxes. Contours of constant induction amplitude ($A$) intersecting Scenarios A, B, and F are shown for the synodic, its second harmonic, and the orbital frequency. Scenarios B, D, and F occupy a region where the synodic induction response is saturated, such that changes in ocean thickness produce negligible changes in the induction response.}
	\label{fig: synodic gradient}
\end{figure}

The second class of failure is represented by the Scenario A Metric II failure for ocean thickness.
Unlike the prior failure, this is not due to the retrieval model but instead stems from the difficulty of breaking the degeneracy between ocean thickness and conductivity in specific regions of parameter space, as evidenced by the parallel contours for the major oscillation frequencies in Figure \ref{fig: synodic gradient}. The resulting broad posterior for the ocean thickness encompasses the input values, but provides only marginal improvement compared to the prior. The same effect causes the ocean thickness failure in Scenario F (Metric II).

In Scenarios D and E, the ocean thickness posteriors are also somewhat broad and uniform, though in these cases the parameter is able to be recovered successfully. The remaining scenarios, C and G, exhibit nearly complete and partial degeneracy breaking, respectively, and all parameters are recovered successfully in both scenarios.

We find no relationship between the magnitude of sensor offsets and drift rates and retrieval performance. For example, Scenarios C and D have the largest and smallest magnitude offsets, respectively, but both retrievals are successful and the only marginal score occurs in Scenario D. An extensive Monte Carlo study of instrumental noise is beyond the scope of this study, but limited experiments re-analyzing the same scenario with new noise draws show little change in the recovered posteriors. Variations in assessed performance in different ocean scenarios therefore reflect properties of the assumed internal structure, and are likely robust to the specifics of instrument noise at the levels considered here.

\subsection{Moon-Plasma Interaction Fields}
\label{sec: mpi retrievals}
For analysis of actual ECM data, it is expected that MHD simulations \citep[e.g.,][]{2020JA028888} will be used to model the magnetic fields generated by the interaction of the magnetospheric plasma with Europa, allowing these confounding signals to be removed from the magnetometry. However, this process is imperfect, leaving residual noise in the data used for induction studies. To investigate this effect, we generated synthetic data sets for each of the scenarios with residual plasma fields reduced by $90\%$ compared to their original intensity (see Section \ref{sec: mpi}).

\begin{figure}
	\centering
	\includegraphics[width=0.5\textwidth]{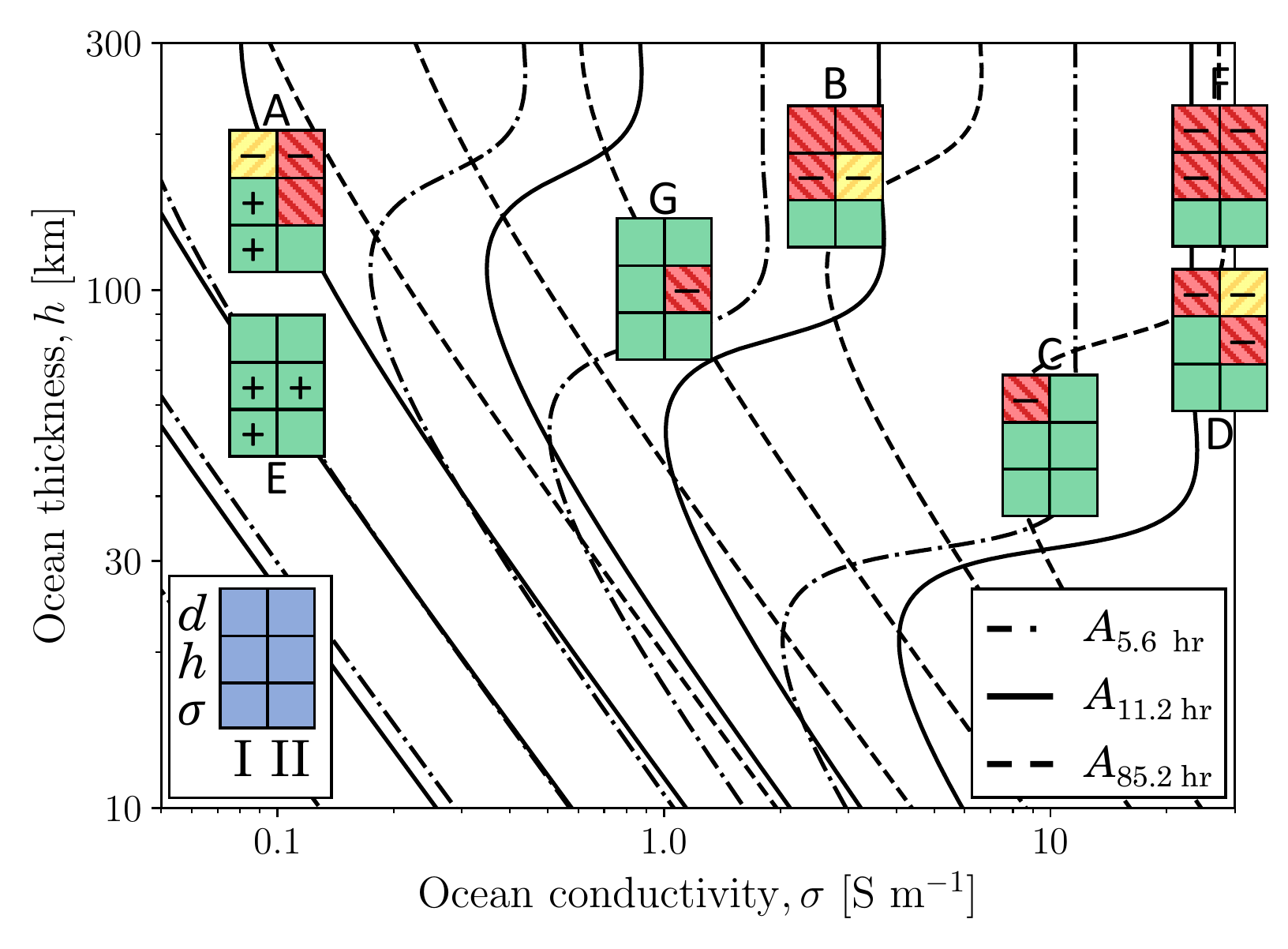}
	\caption{\small Performance of Bayesian retrieval method for all scenarios with incomplete ($90\%$) removal of moon-plasma interaction fields from ECM magnetometry. Solid green boxes correspond to successful recovery of a parameter as measured by the indicated metric. Yellow boxes with positively sloped stripes indicate a marginal recovery and red boxes with negatively sloped stripes indicate a failure. Contours of constant induction response assuming no ice shell ($d=0$) are shown at amplitudes of $25, 50, 75, 90, \text{ and } 95\%$ for the three indicated frequencies and at $98\%$ for the $5.6$ and $11.2~\mathrm{hr}$ periods. Improved or degraded performance grade compared to baseline is indicated with a $+$ or $-$, respectively.}
	\label{fig: plasma10 results}
	% 9 Freq: 26 success, 3 marginal, 13 failure
\end{figure}

\begin{figure*}
	\centering
	\includegraphics[width=\textwidth]{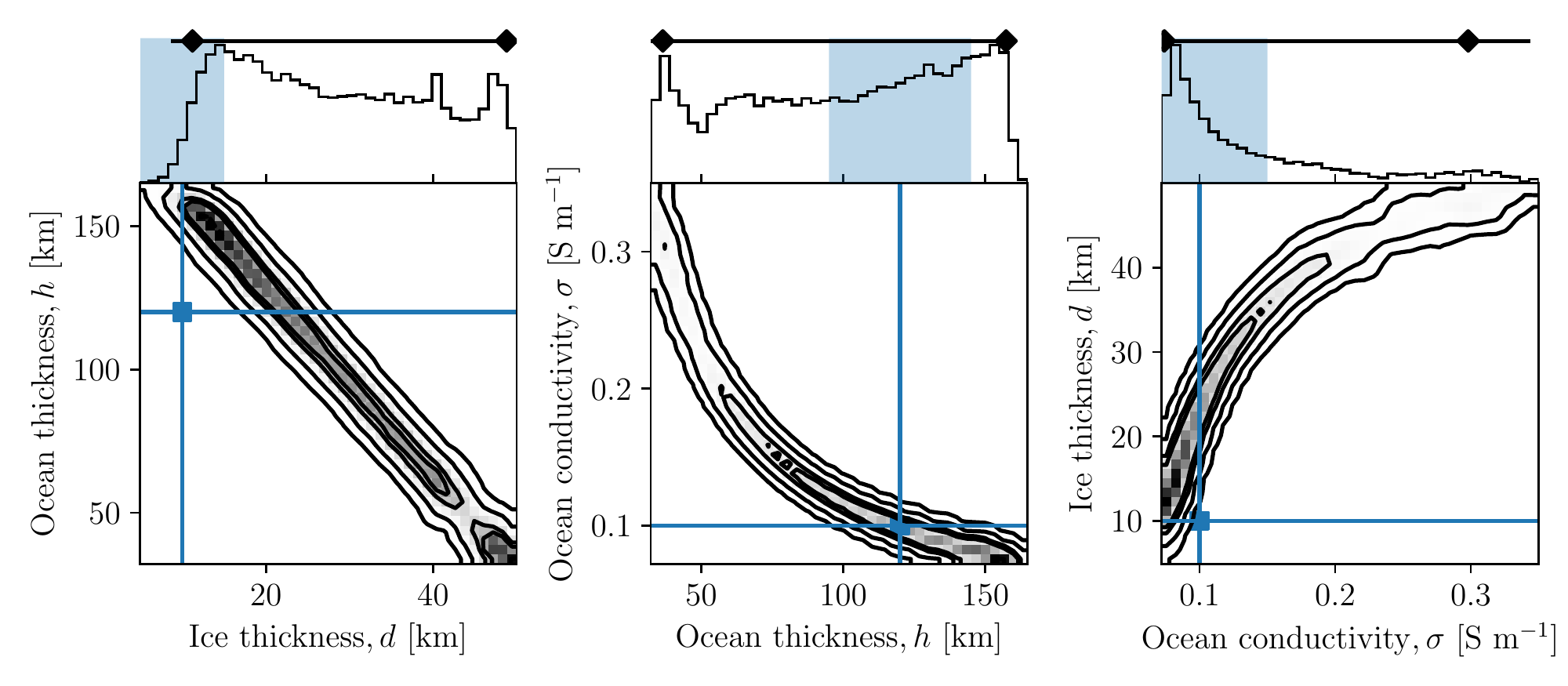}
	\caption{\small Example retrieval of Europan interior for Scenario A assuming partial ($90\%$) removal of moon-plasma interaction fields (compare to Figure \ref{fig: baseline scenario a}) with 1D and 2D posterior distributions for the three ocean parameters and parameter pairings. Two-dimensional posteriors illustrated by contours of constant probability density drawn enclosing $5, 50, 68, 95, \text{ and } 99.7\%$ of the probability mass. Blue lines and square markers indicate the input values. Above each contour plot, a histogram displays the posterior for the ocean parameter on the $x$-axis. In these plots, the horizontal black line indicates the $99.7\%$ interval, the black diamonds indicate the bounds of the $95\%$ interval, and the blue shaded region indicates the Metric II interval.}
	\label{fig: plasma scenario a}
\end{figure*}

We find that despite the relatively large amplitude of the plasma residuals, which are sometimes greater than the oscillation amplitude of the driving field for all but two frequencies, our retrievals are fairly robust to the presence of plasma effects (Figure \ref{fig: plasma10 results}). However, the ability to accurately recover the ice shell thickness, and to a lesser extent the ocean thickness, is somewhat diminished. 
Across four scenarios (A, C, D, F), seven ice shell retrieval metrics show degradation, including five new failures, while in Scenarios B, D, F, and G four new ocean thickness failures occur, with one additional marginal score. There is no significant decrease in our ability to recover the ocean conductivity and, in five cases, marginal scores improve to successes. 

These changes are exemplified in the new recovery of the ocean parameters in Scenario A (Figure \ref{fig: plasma scenario a}). All three posteriors have broadened compared to the case without moon-plasma interaction fields, particularly the recovered ice and ocean thickness distributions which are now nearly uniform and span almost the full range permitted by the priors. In addition to broadening, the ice shell posterior is offset from the input value so that the true value falls only within the $99.7\%$ interval. Combined, these effects convert successful scores on Metrics I and II to marginal and failure, respectively. The ocean thickness remains poorly constrained in this recovery and continues to fail Metric II. However, its broadened posterior more easily captures the input value, resulting in an improvement from marginal to successful in Metric I. Finally, despite the unbroken degeneracy between ocean conductivity and ocean thickness, the conductivity remains well-constrained and is recovered successfully.

Adding the plasma interaction has limited deleterious effects because the associated magnetic fluctuations typically do not match the specific spatial and temporal signature associated with a dipole oscillating at a single, fixed frequency.
Therefore, most of the plasma field cannot be fit by the induction model and is treated as noise. This additional noise causes the recovered posteriors to broaden, resulting in degraded characterization (Metric II; Scenarios A, B, D, F, G). Conversely, these wider credible intervals also increase the likelihood of capturing the input value, improving some scores on Metric I in Scenarios A and E. The small component of the plasma perturbation that is consistent with an induced field, which can therefore be fit by the retrieval model, introduces errors in the recovery, but these errors are unlikely to produce the self-consistent change in induction amplitude and phase lag required to bias the ocean thickness and conductivity. This is illustrated by the recovered joint posterior distribution of ocean thickness and conductivity, which contains the input value in the $50\%$ credible region (Figure \ref{fig: plasma scenario a}, middle panel). The ice shell, by contrast, is principally constrained by the induction amplitude, making it more susceptible to errors from plasma effects (Scenarios A, C, D, F).

\subsection{Informative Priors}
\label{sec: informative priors}
In the preceding analysis, the recovered posteriors are obtained from simulated Europa Clipper magnetometry and loosely-confining priors derived largely from observations taken by Galileo. However, Europa Clipper will fly a suite of instruments that provide complementary insights into Europa's internal structure. In particular, measurements of static gravity by the Gravity/Radio Science investigation \citep[GRS,][]{2021LPI....52.1784M} are expected to constrain the total thickness of the hydrosphere, but to have difficulty determining the location of the ice-ocean boundary owing to the similar densities of frozen and liquid water \citep{1998Sci...281.2019A, 2021Icar..35814187G}. Here we consider how combining these other constraints with magnetometry measurements can enable more accurate and precise inversions for ocean structures. 

\begin{figure}
	\centering
	\includegraphics[width=0.5\textwidth]{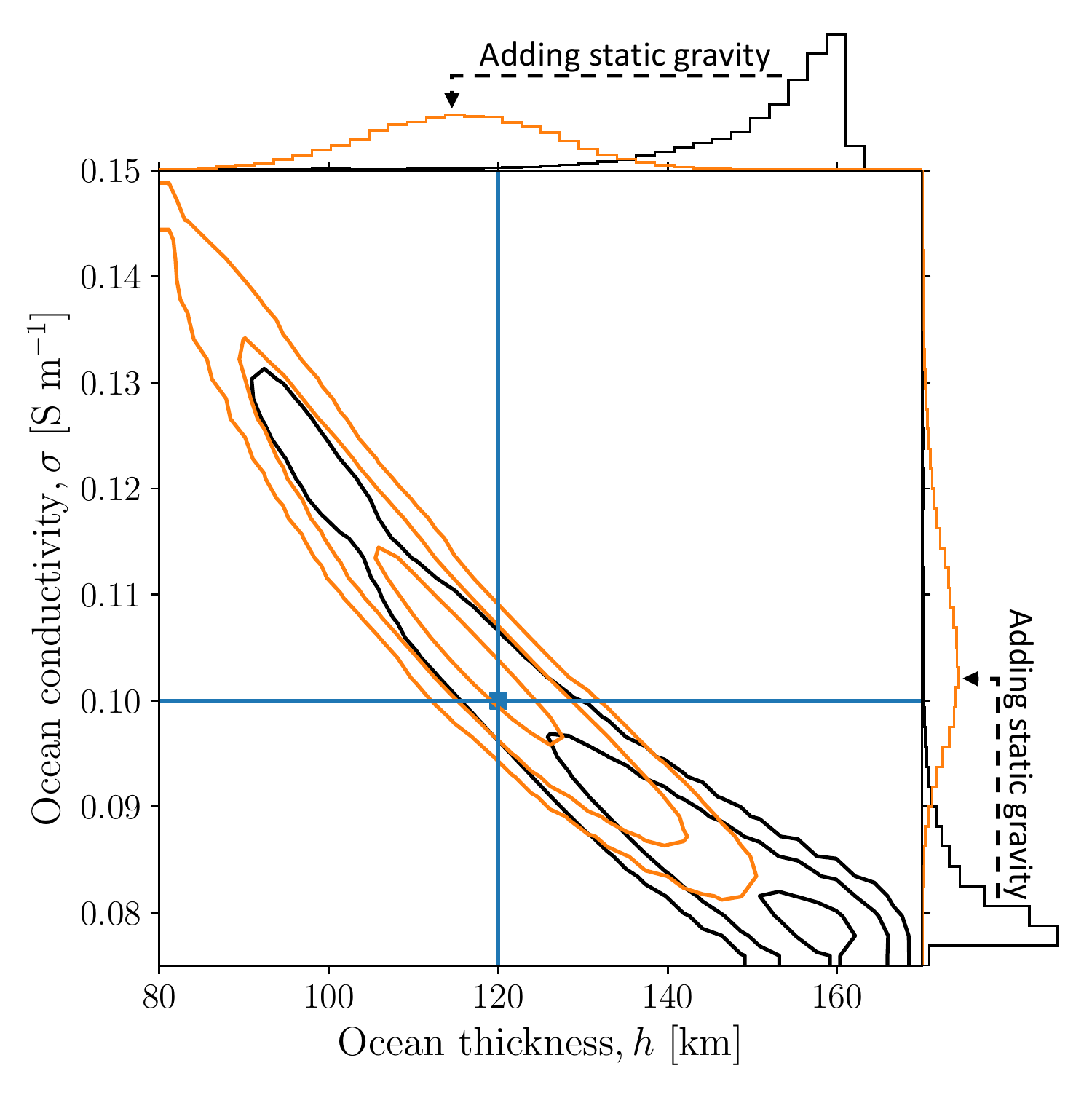}
	\caption{\small Joint magnetic and gravity inversion for Europa's ocean. Shown are recovered posterior distributions when using a gravity-derived prior for the hydrosphere thickness, $d + h \sim \mathcal{N}{(130, 7.5)}$ as compared to a less informative uniform distribution, $d + h \sim \mathcal{U}{(80, 170)}$. A 2D histogram of joint posterior distribution of ocean thickness and conductivity is shown with marginalized 1D histograms on each axis. The posterior using the uniform and normal priors are shown in black and orange, respectively, with contour lines drawn at the $50$th, $95$th, and $99.7$th percentiles. The blue lines mark the input values for the chosen scenario (A). While both posteriors include the correct value, using the gravity-informed prior provides improved ocean characterization.}
	\label{fig: scenario a informative prior}
\end{figure}

To investigate a simple joint recovery of the interior using static gravity and induction we conducted an additional retrieval of Scenario A, this time imposing a more informative prior on the hydrosphere thickness. 
Although the independent recovery of the hydrosphere thickness by GRS has not been fully established \citep[e.g.,][]{PetriccaEGU2022}, we evaluate how an unbiased estimate with a 1-$\sigma$ accuracy of $7.5~\mathrm{km}$ would affect our results.
The uniform prior from $80\text{--}170~\mathrm{km}$ is therefore replaced by a normal distribution, $d + h \sim \mathcal{N}{(130~\mathrm{km}, 7.5~\mathrm{km})}$. We find this improves the ocean retrieval, such that both the ocean conductivity and thickness are successfully recovered (Figure 
\ref{fig: scenario a informative prior}). Notably, the degeneracy between conductivity and ocean thickness remains largely unbroken; as a result the range of values spanned by the posteriors is not substantially changed. Instead, the additional information has pulled the probability mass towards the correct values, yielding a less biased estimation. This example indicates that more sophisticated retrievals that jointly use magnetometry, gravity science, and other Clipper observations are likely to provide the clearest picture of Europa's interior structure.

%%%%% Section: Discussion and conclusion
\section{Discussion}
\label{sec: discussion}
We have developed a novel Bayesian framework for inverting spacecraft magnetometry to recover the internal structure of icy moons and applied it to the upcoming Europa Clipper mission. In contrast with earlier work \citep[e.g.,][]{2004JGRE..109.5006S}, our method uses a three-layer model of the internal structure to generate a self-consistent induction response at multiple frequencies, allowing us to directly recover the ocean parameters and associated uncertainties from the magnetometry. This is particularly useful as we move from ocean detection with Galileo to characterization with Europa Clipper, an advance made possible by the much larger number of Europa flybys Clipper will perform to enable multi-frequency induction studies.

Through the use of realistic simulated data, we demonstrate that the combination of our Bayesian approach and ECM data will be able to significantly advance our understanding of Europa's interior. 
By our metrics, the ocean conductivity can be successfully or marginally recovered across a wide range of plausible Europan interiors and the ice shell thickness is recovered except in one thin shell ($d = 2~\mathrm{km}$) scenario. Accurate recovery of the ocean thickness is more scenario dependent, with some combinations of ocean thickness and conductivity producing degenerate solutions that prevent a unique identification of the ocean thickness. This is naturally incorporated in the estimates for the interior parameters, making our uncertainty estimates robust to model degeneracies. 
The addition of simulated moon-plasma interaction fields \citep[e.g.,][]{2020JA028888} creates considerable artifacts in the magnetic data used in induction analysis. However, we find that our recoveries are surprisingly robust to incomplete ($90\%$) removal of these effects even though the remaining features can have ${\sim}10 ~\mathrm{nT}$ amplitude. Specifically, ocean conductivity estimates are largely unaffected, though the ocean thickness and ice thickness recoveries are significantly degraded. Accurate measurements and MHD modeling of the Europan plasma environment to achieve better than $90\%$ interaction field reduction can therefore improve ocean conductivity estimates and are crucial to accurate ocean thickness and ice shell characterization.

Beyond ECM and PIMS, Europa Clipper will carry a suite of additional instruments that can be used to complement the insights obtained from magnetometry.
For example, we demonstrate that measurements of static gravity can provide useful constraints on the hydrosphere thickness, which improve ocean characterization (Section \ref{sec: informative priors}). 
Sub-surface sounding by Europa Clipper's radar \citep{2009euro.book..631B}, measurements of Europa's tidal deformation \citep[][]{2015GeoRL..42.3166M, 2018Icar..314...35V}, and improved understanding of the thermal constraints on Europa's ice shell \citep[e.g.,][]{2021PSJ.....2..129H} may all similarly offer means to improve the recovery of Europa's interior, either through joint analysis or construction of appropriate priors. 

Recent work has explored the magnetic signature produced by more complex internal structure models, including non-spherical oceans \citep{2021Icar..35414020S}, as well as radially varying conductivity and induced fields produced by ocean flows \citep{2021JGRE..12606418V}. The error introduced by the assumption of spherical symmetry is expected to be small, ${\sim}0.5 ~\mathrm{nT}$ at Europa's surface \citep{2021Icar..35414020S}. Similarly, among the scenarios considered in \citet{2021JGRE..12606418V}, the induction response from a Europan ocean with a self-consistent adiabatic profile and radially varying conductivity differs from a uniform conductivity ocean with the same mean conductivity by ${\lesssim}1 ~\mathrm{nT}$ at the surface. Motional induction created by ocean currents can plausibly produce much larger signatures (${\lesssim}20 ~\mathrm{nT}$), but this is contingent on the flow velocity, which is uncertain \citep{2021JGRE..12606418V}. While these effects may be detectable on some flybys, because of their expected scale compared to the noise sources included in this analysis, we do not expect them to significantly change our estimate of ECM's ability to recover the global properties of the ocean.

Building a more granular picture of Europa's ocean, however, will require incorporating these more sophisticated internal structure models into our Bayesian framework. In particular, ongoing work includes adopting models with radial conductivity structure \citep{1963Eckhardt, 1966SrivastavaSphericalMulti} to both probe the ocean's thermal structure, composition, and dynamics \citep{2021JGRE..12606418V} and enable disentangling ionospheric induction from the oceanic signal \citep[e.g.,][]{2022E&SS....902034C}. Future work will apply these expanded capabilities to archival data and, eventually, magnetometry from Europa Clipper. As our exploration of ocean worlds in the solar system continues, Bayesian inference will offer a flexible and powerful tool for magnetic induction investigations to help reveal the interiors of these icy worlds.

\begin{acknowledgments}
JBB and BPW thank the NASA Europa Clipper project (University of Michigan/JPL SUBK00011438) and the MIT Dean of Science Discretionary Fund for support. JBB thanks C. Carroll and K. Knudson for insightful discussions. BPW thanks Ben Vigoda for thoughtful discussions about the meaning and power of Bayesian inference.
\end{acknowledgments}

\software{ArviZ \citep{arviz_2019}, Astropy \citep{2013A&A...558A..33A,2018AJ....156..123A}, emcee \citep{2013PASP..125..306F}, matplotlib \citep{Hunter:2007}, numpy \citep{harris2020array}, PyMC3 \citep{2016ascl.soft10016S}, SciPy \citep{2020SciPy-NMeth}}

\bibliographystyle{aasjournal}
\bibliography{EuropaBayesReferences}

\begin{thebibliography}{}
\expandafter\ifx\csname natexlab\endcsname\relax\def\natexlab#1{#1}\fi
\providecommand{\url}[1]{\href{#1}{#1}}
\providecommand{\dodoi}[1]{doi:~\href{http://doi.org/#1}{\nolinkurl{#1}}}
\providecommand{\doeprint}[1]{\href{http://ascl.net/#1}{\nolinkurl{http://ascl.net/#1}}}
\providecommand{\doarXiv}[1]{\href{https://arxiv.org/abs/#1}{\nolinkurl{https://arxiv.org/abs/#1}}}

\bibitem[{{Acton} {et~al.}(2018){Acton}, {Bachman}, {Semenov}, \&
  {Wright}}]{2018P&SS..150....9A}
{Acton}, C., {Bachman}, N., {Semenov}, B., \& {Wright}, E. 2018, \planss, 150,
  9, \dodoi{10.1016/j.pss.2017.02.013}

\bibitem[{{Anderson} {et~al.}(1998){Anderson}, {Schubert}, {Jacobson}, {Lau},
  {Moore}, \& {Sjogren}}]{1998Sci...281.2019A}
{Anderson}, J.~D., {Schubert}, G., {Jacobson}, R.~A., {et~al.} 1998, Science,
  281, 2019, \dodoi{10.1126/science.281.5385.2019}

\bibitem[{{Archinal} {et~al.}(2018){Archinal}, {Acton}, {A'Hearn}, {Conrad},
  {Consolmagno}, {Duxbury}, {Hestroffer}, {Hilton}, {Kirk}, {Klioner},
  {McCarthy}, {Meech}, {Oberst}, {Ping}, {Seidelmann}, {Tholen}, {Thomas}, \&
  {Williams}}]{2018CeMDA.130...22A}
{Archinal}, B.~A., {Acton}, C.~H., {A'Hearn}, M.~F., {et~al.} 2018, Celestial
  Mechanics and Dynamical Astronomy, 130, 22, \dodoi{10.1007/s10569-017-9805-5}

\bibitem[{{Astropy Collaboration} {et~al.}(2013){Astropy Collaboration},
  {Robitaille}, {Tollerud}, {Greenfield}, {Droettboom}, {Bray}, {Aldcroft},
  {Davis}, {Ginsburg}, {Price-Whelan}, {Kerzendorf}, {Conley}, {Crighton},
  {Barbary}, {Muna}, {Ferguson}, {Grollier}, {Parikh}, {Nair}, {Unther},
  {Deil}, {Woillez}, {Conseil}, {Kramer}, {Turner}, {Singer}, {Fox}, {Weaver},
  {Zabalza}, {Edwards}, {Azalee Bostroem}, {Burke}, {Casey}, {Crawford},
  {Dencheva}, {Ely}, {Jenness}, {Labrie}, {Lim}, {Pierfederici}, {Pontzen},
  {Ptak}, {Refsdal}, {Servillat}, \& {Streicher}}]{2013A&A...558A..33A}
{Astropy Collaboration}, {Robitaille}, T.~P., {Tollerud}, E.~J., {et~al.} 2013,
  \aap, 558, A33, \dodoi{10.1051/0004-6361/201322068}

\bibitem[{{Astropy Collaboration} {et~al.}(2018){Astropy Collaboration},
  {Price-Whelan}, {Sip{\H{o}}cz}, {G{\"u}nther}, {Lim}, {Crawford}, {Conseil},
  {Shupe}, {Craig}, {Dencheva}, {Ginsburg}, {VanderPlas}, {Bradley},
  {P{\'e}rez-Su{\'a}rez}, {de Val-Borro}, {Aldcroft}, {Cruz}, {Robitaille},
  {Tollerud}, {Ardelean}, {Babej}, {Bach}, {Bachetti}, {Bakanov}, {Bamford},
  {Barentsen}, {Barmby}, {Baumbach}, {Berry}, {Biscani}, {Boquien}, {Bostroem},
  {Bouma}, {Brammer}, {Bray}, {Breytenbach}, {Buddelmeijer}, {Burke},
  {Calderone}, {Cano Rodr{\'\i}guez}, {Cara}, {Cardoso}, {Cheedella}, {Copin},
  {Corrales}, {Crichton}, {D'Avella}, {Deil}, {Depagne}, {Dietrich}, {Donath},
  {Droettboom}, {Earl}, {Erben}, {Fabbro}, {Ferreira}, {Finethy}, {Fox},
  {Garrison}, {Gibbons}, {Goldstein}, {Gommers}, {Greco}, {Greenfield},
  {Groener}, {Grollier}, {Hagen}, {Hirst}, {Homeier}, {Horton}, {Hosseinzadeh},
  {Hu}, {Hunkeler}, {Ivezi{\'c}}, {Jain}, {Jenness}, {Kanarek}, {Kendrew},
  {Kern}, {Kerzendorf}, {Khvalko}, {King}, {Kirkby}, {Kulkarni}, {Kumar},
  {Lee}, {Lenz}, {Littlefair}, {Ma}, {Macleod}, {Mastropietro}, {McCully},
  {Montagnac}, {Morris}, {Mueller}, {Mumford}, {Muna}, {Murphy}, {Nelson},
  {Nguyen}, {Ninan}, {N{\"o}the}, {Ogaz}, {Oh}, {Parejko}, {Parley}, {Pascual},
  {Patil}, {Patil}, {Plunkett}, {Prochaska}, {Rastogi}, {Reddy Janga},
  {Sabater}, {Sakurikar}, {Seifert}, {Sherbert}, {Sherwood-Taylor}, {Shih},
  {Sick}, {Silbiger}, {Singanamalla}, {Singer}, {Sladen}, {Sooley},
  {Sornarajah}, {Streicher}, {Teuben}, {Thomas}, {Tremblay}, {Turner},
  {Terr{\'o}n}, {van Kerkwijk}, {de la Vega}, {Watkins}, {Weaver}, {Whitmore},
  {Woillez}, {Zabalza}, \& {Astropy Contributors}}]{2018AJ....156..123A}
{Astropy Collaboration}, {Price-Whelan}, A.~M., {Sip{\H{o}}cz}, B.~M., {et~al.}
  2018, \aj, 156, 123, \dodoi{10.3847/1538-3881/aabc4f}

\bibitem[{{Bagenal} \& {Dols}(2020)}]{2020JGRA..12527485B}
{Bagenal}, F., \& {Dols}, V. 2020, Journal of Geophysical Research (Space
  Physics), 125, e27485, \dodoi{10.1029/2019JA027485}

\bibitem[{{Bagenal} {et~al.}(2015){Bagenal}, {Sidrow}, {Wilson}, {Cassidy},
  {Dols}, {Crary}, {Steffl}, {Delamere}, {Kurth}, \&
  {Paterson}}]{2015Icar..261....1B}
{Bagenal}, F., {Sidrow}, E., {Wilson}, R.~J., {et~al.} 2015, \icarus, 261, 1,
  \dodoi{10.1016/j.icarus.2015.07.036}

\bibitem[{{Billings} \& {Kattenhorn}(2005)}]{2005Icar..177..397B}
{Billings}, S.~E., \& {Kattenhorn}, S.~A. 2005, \icarus, 177, 397,
  \dodoi{10.1016/j.icarus.2005.03.013}

\bibitem[{{Blankenship} {et~al.}(2009){Blankenship}, {Young}, {Moore}, \&
  {Moore}}]{2009euro.book..631B}
{Blankenship}, D.~D., {Young}, D.~A., {Moore}, W.~B., \& {Moore}, J.~C. 2009,
  {Radar Sounding of Europa's Subsurface Properties and Processes: The View
  from Earth}, ed. R.~T. {Pappalardo}, W.~B. {McKinnon}, \& K.~K. {Khurana},
  631

\bibitem[{{Bromund} {et~al.}(2016){Bromund}, {Plaschke}, {Strangeway},
  {Anderson}, {Huang}, {Magnes}, {Fischer}, {Nakamura}, {Leinweber}, {Russell},
  {Baumjohann}, {Chutter}, {Torbert}, {Le}, {Slavin}, \&
  {Kepko}}]{2016AGUFMSM21A2455B}
{Bromund}, K.~R., {Plaschke}, F., {Strangeway}, R.~J., {et~al.} 2016, in AGU
  Fall Meeting Abstracts, SM21A--2455

\bibitem[{{Cochrane} {et~al.}(2022){Cochrane}, {Persinger}, {Vance}, {Midkiff},
  {Castillo-Rogez}, {Luspay-Kuti}, {Liuzzo}, {Paty}, {Mitchell}, \&
  {Prockter}}]{2022E&SS....902034C}
{Cochrane}, C.~J., {Persinger}, R.~R., {Vance}, S.~D., {et~al.} 2022, Earth and
  Space Science, 9, e02034, \dodoi{10.1029/2021EA002034}

\bibitem[{{Connerney} {et~al.}(1998){Connerney}, {Acu{\~n}a}, {Ness}, \&
  {Satoh}}]{1998JGR...10311929C}
{Connerney}, J.~E.~P., {Acu{\~n}a}, M.~H., {Ness}, N.~F., \& {Satoh}, T. 1998,
  \jgr, 103, 11929, \dodoi{10.1029/97JA03726}

\bibitem[{{Connerney} {et~al.}(2018){Connerney}, {Kotsiaros}, {Oliversen},
  {Espley}, {Joergensen}, {Joergensen}, {Merayo}, {Herceg}, {Bloxham}, {Moore},
  {Bolton}, \& {Levin}}]{2018GeoRL..45.2590C}
{Connerney}, J.~E.~P., {Kotsiaros}, S., {Oliversen}, R.~J., {et~al.} 2018,
  \grl, 45, 2590, \dodoi{10.1002/2018GL077312}

\bibitem[{{Dougherty} {et~al.}(2004){Dougherty}, {Kellock}, {Southwood},
  {Balogh}, {Smith}, {Tsurutani}, {Gerlach}, {Glassmeier}, {Gleim}, {Russell},
  {Erdos}, {Neubauer}, \& {Cowley}}]{2004SSRv..114..331D}
{Dougherty}, M.~K., {Kellock}, S., {Southwood}, D.~J., {et~al.} 2004, \ssr,
  114, 331, \dodoi{10.1007/s11214-004-1432-2}

\bibitem[{Eckhardt(1963)}]{1963Eckhardt}
Eckhardt, D.~H. 1963, Journal of Geophysical Research (1896-1977), 68, 6273,
  \dodoi{https://doi.org/10.1029/JZ068i023p06273}

\bibitem[{{Foreman-Mackey} {et~al.}(2013){Foreman-Mackey}, {Hogg}, {Lang}, \&
  {Goodman}}]{2013PASP..125..306F}
{Foreman-Mackey}, D., {Hogg}, D.~W., {Lang}, D., \& {Goodman}, J. 2013, \pasp,
  125, 306, \dodoi{10.1086/670067}

\bibitem[{{Gomez Casajus} {et~al.}(2021){Gomez Casajus}, {Zannoni}, {Modenini},
  {Tortora}, {Nimmo}, {Van Hoolst}, {Buccino}, \&
  {Oudrhiri}}]{2021Icar..35814187G}
{Gomez Casajus}, L., {Zannoni}, M., {Modenini}, D., {et~al.} 2021, \icarus,
  358, 114187, \dodoi{10.1016/j.icarus.2020.114187}

\bibitem[{{Goodman} \& {Weare}(2010)}]{2010CAMCS...5...65G}
{Goodman}, J., \& {Weare}, J. 2010, Communications in Applied Mathematics and
  Computational Science, 5, 65, \dodoi{10.2140/camcos.2010.5.65}

\bibitem[{{Hand} {et~al.}(2007){Hand}, {Carlson}, \&
  {Chyba}}]{2007AsBio...7.1006H}
{Hand}, K.~P., {Carlson}, R.~W., \& {Chyba}, C.~F. 2007, Astrobiology, 7, 1006,
  \dodoi{10.1089/ast.2007.0156}

\bibitem[{{Hand} \& {Chyba}(2007)}]{2007Icar..189..424H}
{Hand}, K.~P., \& {Chyba}, C.~F. 2007, \icarus, 189, 424,
  \dodoi{10.1016/j.icarus.2007.02.002}

\bibitem[{{Hand} {et~al.}(2020){Hand}, {Sotin}, {Hayes}, \&
  {Coustenis}}]{2020SSRv..216...95H}
{Hand}, K.~P., {Sotin}, C., {Hayes}, A., \& {Coustenis}, A. 2020, \ssr, 216,
  95, \dodoi{10.1007/s11214-020-00713-7}

\bibitem[{Harris {et~al.}(2021)Harris, Jia, Slavin, Toth, Huang, \&
  Rubin}]{2020JA028888}
Harris, C. D.~K., Jia, X., Slavin, J.~A., {et~al.} 2021, Journal of Geophysical
  Research: Space Physics, 126, e2020JA028888,
  \dodoi{https://doi.org/10.1029/2020JA028888}

\bibitem[{Harris {et~al.}(2020)Harris, Millman, van~der Walt, Gommers,
  Virtanen, Cournapeau, Wieser, Taylor, Berg, Smith, Kern, Picus, Hoyer, van
  Kerkwijk, Brett, Haldane, del R{\'{i}}o, Wiebe, Peterson,
  G{\'{e}}rard-Marchant, Sheppard, Reddy, Weckesser, Abbasi, Gohlke, \&
  Oliphant}]{harris2020array}
Harris, C.~R., Millman, K.~J., van~der Walt, S.~J., {et~al.} 2020, Nature, 585,
  357, \dodoi{10.1038/s41586-020-2649-2}

\bibitem[{{Howell}(2021)}]{2021PSJ.....2..129H}
{Howell}, S.~M. 2021, \psj, 2, 129, \dodoi{10.3847/PSJ/abfe10}

\bibitem[{{Howell} \& {Pappalardo}(2020)}]{2020NatCo..11.1311H}
{Howell}, S.~M., \& {Pappalardo}, R.~T. 2020, Nature Communications, 11, 1311,
  \dodoi{10.1038/s41467-020-15160-9}

\bibitem[{Hunter(2007)}]{Hunter:2007}
Hunter, J.~D. 2007, Computing in Science \& Engineering, 9, 90,
  \dodoi{10.1109/MCSE.2007.55}

\bibitem[{Jackson {et~al.}(2020)Jackson, Bayer, Sheldon, Buffington, Harris,
  Jones-Wilson, Laslo, Lee, Perez, Salami, Schimmels, Smith, Soriano, \&
  Wang}]{9172447}
Jackson, M., Bayer, T., Sheldon, C., {et~al.} 2020, in 2020 IEEE Aerospace
  Conference, 1--20, \dodoi{10.1109/AERO47225.2020.9172447}

\bibitem[{{Khurana}(1997)}]{1997JGR...10211295K}
{Khurana}, K.~K. 1997, \jgr, 102, 11295, \dodoi{10.1029/97JA00563}

\bibitem[{{Khurana}(2001)}]{2001JGR...10625999K}
---. 2001, \jgr, 106, 25999, \dodoi{10.1029/2000JA000352}

\bibitem[{{Khurana} {et~al.}(2009){Khurana}, {Kivelson}, {Hand}, \&
  {Russell}}]{2009euro.book..571K}
{Khurana}, K.~K., {Kivelson}, M.~G., {Hand}, K.~P., \& {Russell}, C.~T. 2009,
  {Electromagnetic Induction from Europa's Ocean and the Deep Interior}, ed.
  R.~T. {Pappalardo}, W.~B. {McKinnon}, \& K.~K. {Khurana}, 571

\bibitem[{{Khurana} {et~al.}(2002){Khurana}, {Kivelson}, \&
  {Russell}}]{2002AsBio...2...93K}
{Khurana}, K.~K., {Kivelson}, M.~G., \& {Russell}, C.~T. 2002, Astrobiology, 2,
  93, \dodoi{10.1089/153110702753621376}

\bibitem[{{Khurana} {et~al.}(1998){Khurana}, {Kivelson}, {Stevenson},
  {Schubert}, {Russell}, {Walker}, \& {Polanskey}}]{1998Natur.395..777K}
{Khurana}, K.~K., {Kivelson}, M.~G., {Stevenson}, D.~J., {et~al.} 1998, \nat,
  395, 777, \dodoi{10.1038/27394}

\bibitem[{{Khurana} \& {Schwarzl}(2005)}]{2005JGRA..110.7227K}
{Khurana}, K.~K., \& {Schwarzl}, H.~K. 2005, Journal of Geophysical Research
  (Space Physics), 110, A07227, \dodoi{10.1029/2004JA010757}

\bibitem[{{Khurana} {et~al.}(2004){Khurana}, {Tsyganenko}, \&
  {Schwartzl}}]{2004cosp...35.2073K}
{Khurana}, K.~K., {Tsyganenko}, N.~A., \& {Schwartzl}, H.~K. 2004, in 35th
  COSPAR Scientific Assembly, Vol.~35, 2073

\bibitem[{{Kivelson} {et~al.}(2000){Kivelson}, {Khurana}, {Russell}, {Volwerk},
  {Walker}, \& {Zimmer}}]{2000Sci...289.1340K}
{Kivelson}, M.~G., {Khurana}, K.~K., {Russell}, C.~T., {et~al.} 2000, Science,
  289, 1340, \dodoi{10.1126/science.289.5483.1340}

\bibitem[{{Kivelson} {et~al.}(2009){Kivelson}, {Khurana}, \&
  {Volwerk}}]{2009euro.book..545K}
{Kivelson}, M.~G., {Khurana}, K.~K., \& {Volwerk}, M. 2009, {Europa's
  Interaction with the Jovian Magnetosphere}, ed. R.~T. {Pappalardo}, W.~B.
  {McKinnon}, \& K.~K. {Khurana}, 545

\bibitem[{Kumar {et~al.}(2019)Kumar, Carroll, Hartikainen, \&
  Martin}]{arviz_2019}
Kumar, R., Carroll, C., Hartikainen, A., \& Martin, O. 2019, Journal of Open
  Source Software, 4, 1143, \dodoi{10.21105/joss.01143}

\bibitem[{{Mazarico} {et~al.}(2015){Mazarico}, {Genova}, {Neumann}, {Smith}, \&
  {Zuber}}]{2015GeoRL..42.3166M}
{Mazarico}, E., {Genova}, A., {Neumann}, G.~A., {Smith}, D.~E., \& {Zuber},
  M.~T. 2015, \grl, 42, 3166, \dodoi{10.1002/2015GL063224}

\bibitem[{{Mazarico} {et~al.}(2021){Mazarico}, {Buccino}, {Castillo-Rogez},
  {Dombard}, {Genova}, {Hussmann}, {Kiefer}, {Lunine}, {McKinnon}, {Nimmo},
  {Park}, {Tortora}, {Withers}, {Roberts}, {Korth}, {Senske}, \&
  {Pappalardo}}]{2021LPI....52.1784M}
{Mazarico}, E., {Buccino}, D.~R., {Castillo-Rogez}, J., {et~al.} 2021, in 52nd
  Lunar and Planetary Science Conference, Lunar and Planetary Science
  Conference, 1784

\bibitem[{McKay {et~al.}(1979)McKay, Beckman, \& Conover}]{10.2307/1268522}
McKay, M.~D., Beckman, R.~J., \& Conover, W.~J. 1979, Technometrics, 21, 239.
\newblock \url{http://www.jstor.org/stable/1268522}

\bibitem[{{Neubauer}(1998)}]{1998JGR...10319843N}
{Neubauer}, F.~M. 1998, \jgr, 103, 19843, \dodoi{10.1029/97JE03370}

\bibitem[{{Nimmo} \& {Pappalardo}(2016)}]{2016JGRE..121.1378N}
{Nimmo}, F., \& {Pappalardo}, R.~T. 2016, Journal of Geophysical Research
  (Planets), 121, 1378, \dodoi{10.1002/2016JE005081}

\bibitem[{Parkinson(1983)}]{osti_6997191}
Parkinson, W.~D. 1983, Introduction to geomagnetism (Edinburgh: Scottish
  Academic Press), 308--340

\bibitem[{{Petricca} {et~al.}(2022){Petricca}, {Genova}, {Castillo-Rogez}, \&
  {Mazarico}}]{PetriccaEGU2022}
{Petricca}, F., {Genova}, A., {Castillo-Rogez}, J., \& {Mazarico}, E. 2022, in
  EGU General Assembly 2022, European Geosciences Union No. 5851,
  \dodoi{10.5194/egusphere-egu22-5851}

\bibitem[{{Raymond} {et~al.}(2015){Raymond}, {Jia}, {Joy}, {Khurana}, {Murphy},
  {Russell}, {Strangeway}, \& {Weiss}}]{2015AGUFM.P13E..08R}
{Raymond}, C.~A., {Jia}, X., {Joy}, S.~P., {et~al.} 2015, in AGU Fall Meeting
  Abstracts, Vol. 2015, P13E--08

\bibitem[{{Russell} {et~al.}(2016){Russell}, {Anderson}, {Baumjohann},
  {Bromund}, {Dearborn}, {Fischer}, {Le}, {Leinweber}, {Leneman}, {Magnes},
  {Means}, {Moldwin}, {Nakamura}, {Pierce}, {Plaschke}, {Rowe}, {Slavin},
  {Strangeway}, {Torbert}, {Hagen}, {Jernej}, {Valavanoglou}, \&
  {Richter}}]{2016SSRv..199..189R}
{Russell}, C.~T., {Anderson}, B.~J., {Baumjohann}, W., {et~al.} 2016, \ssr,
  199, 189, \dodoi{10.1007/s11214-014-0057-3}

\bibitem[{{Salvatier} {et~al.}(2016){Salvatier}, {Wiecki}, \&
  {Fonnesbeck}}]{2016ascl.soft10016S}
{Salvatier}, J., {Wiecki}, T.~V., \& {Fonnesbeck}, C. 2016, {PyMC3: Python
  probabilistic programming framework}, Astrophysics Source Code Library,
  record ascl:1610.016.
\newblock \doeprint{1610.016}

\bibitem[{{Saur} {et~al.}(2010){Saur}, {Neubauer}, \&
  {Glassmeier}}]{2010SSRv..152..391S}
{Saur}, J., {Neubauer}, F.~M., \& {Glassmeier}, K.-H. 2010, \ssr, 152, 391,
  \dodoi{10.1007/s11214-009-9581-y}

\bibitem[{{Schilling} {et~al.}(2004){Schilling}, {Khurana}, \&
  {Kivelson}}]{2004JGRE..109.5006S}
{Schilling}, N., {Khurana}, K.~K., \& {Kivelson}, M.~G. 2004, Journal of
  Geophysical Research (Planets), 109, E05006, \dodoi{10.1029/2003JE002166}

\bibitem[{{Schilling} {et~al.}(2007){Schilling}, {Neubauer}, \&
  {Saur}}]{2007Icar..192...41S}
{Schilling}, N., {Neubauer}, F.~M., \& {Saur}, J. 2007, \icarus, 192, 41,
  \dodoi{10.1016/j.icarus.2007.06.024}

\bibitem[{{Seufert} {et~al.}(2011){Seufert}, {Saur}, \&
  {Neubauer}}]{2011Icar..214..477S}
{Seufert}, M., {Saur}, J., \& {Neubauer}, F.~M. 2011, \icarus, 214, 477,
  \dodoi{10.1016/j.icarus.2011.03.017}

\bibitem[{{Soderlund} {et~al.}(2020){Soderlund}, {Kalousov{\'a}}, {Buffo},
  {Glein}, {Goodman}, {Mitri}, {Patterson}, {Postberg}, {Rovira-Navarro},
  {R{\"u}ckriemen}, {Saur}, {Schmidt}, {Sotin}, {Spohn}, {Tobie}, {Van Hoolst},
  {Vance}, \& {Vermeersen}}]{2020SSRv..216...80S}
{Soderlund}, K.~M., {Kalousov{\'a}}, K., {Buffo}, J.~J., {et~al.} 2020, \ssr,
  216, 80, \dodoi{10.1007/s11214-020-00706-6}

\bibitem[{Srivastava(1966)}]{1966SrivastavaSphericalMulti}
Srivastava, S.~P. 1966, Geophysical Journal International, 11, 373,
  \dodoi{10.1111/j.1365-246X.1966.tb03090.x}

\bibitem[{{Styczinski} \& {Harnett}(2021)}]{2021Icar..35414020S}
{Styczinski}, M.~J., \& {Harnett}, E.~M. 2021, \icarus, 354, 114020,
  \dodoi{10.1016/j.icarus.2020.114020}

\bibitem[{ter Braak \& Vrugt(2008)}]{terBraak2008SaC}
ter Braak, C. J.~F., \& Vrugt, J.~A. 2008, Statistics and Computing, 18, 435,
  \dodoi{10.1007/s11222-008-9104-9}

\bibitem[{{Vance} {et~al.}(2016){Vance}, {Hand}, \&
  {Pappalardo}}]{2016GeoRL..43.4871V}
{Vance}, S.~D., {Hand}, K.~P., \& {Pappalardo}, R.~T. 2016, \grl, 43, 4871,
  \dodoi{10.1002/2016GL068547}

\bibitem[{{Vance} {et~al.}(2021){Vance}, {Styczinski}, {Bills}, {Cochrane},
  {Soderlund}, {G{\'o}mez-P{\'e}rez}, \& {Paty}}]{2021JGRE..12606418V}
{Vance}, S.~D., {Styczinski}, M.~J., {Bills}, B.~G., {et~al.} 2021, Journal of
  Geophysical Research (Planets), 126, e06418, \dodoi{10.1029/2020JE006418}

\bibitem[{{Verma} \& {Margot}(2018)}]{2018Icar..314...35V}
{Verma}, A.~K., \& {Margot}, J.-L. 2018, \icarus, 314, 35,
  \dodoi{10.1016/j.icarus.2018.05.018}

\bibitem[{Virtanen {et~al.}(2020)Virtanen, Gommers, Oliphant, Haberland, Reddy,
  Cournapeau, Burovski, Peterson, Weckesser, Bright, {van der Walt}, Brett,
  Wilson, Millman, Mayorov, Nelson, Jones, Kern, Larson, Carey, Polat, Feng,
  Moore, {VanderPlas}, Laxalde, Perktold, Cimrman, Henriksen, Quintero, Harris,
  Archibald, Ribeiro, Pedregosa, {van Mulbregt}, \& {SciPy 1.0
  Contributors}}]{2020SciPy-NMeth}
Virtanen, P., Gommers, R., Oliphant, T.~E., {et~al.} 2020, Nature Methods, 17,
  261, \dodoi{10.1038/s41592-019-0686-2}

\bibitem[{{Zimmer} {et~al.}(2000){Zimmer}, {Khurana}, \&
  {Kivelson}}]{2000Icar..147..329Z}
{Zimmer}, C., {Khurana}, K.~K., \& {Kivelson}, M.~G. 2000, \icarus, 147, 329,
  \dodoi{10.1006/icar.2000.6456}

\end{thebibliography}

\end{document}